\documentclass[11pt,a4paper]{article}
\usepackage{times,latexsym}
\usepackage{url}
\usepackage[T1]{fontenc}
\usepackage{CJKutf8}

   \usepackage[acceptedWithA]{tacl2021v1}
\usepackage{tacl2021v1}

\usepackage{xspace,mfirstuc,tabulary}

\newif\iftaclinstructions
\taclinstructionsfalse 
\iftaclinstructions
\renewcommand{\confidential}{}
\renewcommand{\anonsubtext}{(No author info supplied here, for consistency with
TACL-submission anonymization requirements)}
\newcommand{\instr}
\fi

\iftaclpubformat 

\else

\fi

\title{Beyond Theoretical Bounds: Empirical Privacy Loss Calibration \\for Text Rewriting Under Local Differential Privacy}

\author{
  Weijun Li\textsuperscript{1},
  Arnaud Grivet Sébert\textsuperscript{1},
  Qiongkai Xu\textsuperscript{1},
  Annabelle McIver\textsuperscript{1},
  Mark Dras\textsuperscript{1} \\
  \textsuperscript{1}School of Computing, FSE, Macquarie University, Sydney, Australia \\
  \texttt{weijun.li1@hdr.mq.edu.au, arnaud.grivetsebert@gmail.com} \\
  \texttt{\{qiongkai.xu, annabelle.mciver, mark.dras\}@mq.edu.au}
}

\date{}

\usepackage{amsmath,amsfonts,bm}

\def\eqref#1{equation~\ref{#1}}

\def\1{\bm{1}}

\DeclareMathAlphabet{\mathsfit}{\encodingdefault}{\sfdefault}{m}{sl}
\SetMathAlphabet{\mathsfit}{bold}{\encodingdefault}{\sfdefault}{bx}{n}

\usepackage{stmaryrd}
\usepackage{amsfonts}
\usepackage{amssymb}
\usepackage{amsmath}
\usepackage{mathtools}
\usepackage{bbm}
\usepackage{xspace}
\usepackage{todonotes}
\usepackage{algorithm}
\usepackage{algpseudocode}
\usepackage[most]{tcolorbox}
\usepackage{soul} %

\usepackage{wasysym}  %

\usepackage{tabularx,booktabs,multirow}  %
\usepackage{makecell}
\usepackage{subcaption}
\usepackage{enumerate}
\usepackage{siunitx}
\usepackage{wrapfig}  %

\usepackage{listings}
\usepackage{xcolor}  %

\lstdefinelanguage{json}{
    basicstyle=\ttfamily\small,
    showstringspaces=false,
    breaklines=true,
    frame=single,
    backgroundcolor=\color{gray!10}, %
    keywordstyle=\bfseries\color{blue}, %
    stringstyle=\color{teal}, %
    morestring=[b]",
    morecomment=[l]{//},
    morecomment=[s]{/*}{*/},
    morekeywords={true,false,null} %
}

\usepackage{enumitem}
\setlist{leftmargin=*,nosep,topsep=2pt}

\newcommand{\updates}[1]{\textcolor{black}{#1}}
\definecolor{refblue}{RGB}{90,110,130}
\definecolor{lightblue}{HTML}{5E93CC}

\newcommand{\ourmethod}{\text{TeDA}\xspace}
\newcommand{\ourmethodname}{\text{TeDA}\xspace}

\newcommand{\ourmethodfullbold}{\text{\textbf{Te}xt \textbf{D}istinguishability \textbf{A}udit}}\xspace
\newcommand{\ourmethodfull}{\text{Text Distinguishability Audit}\xspace}

\usepackage{ntheorem}

\theoremstyle{nonumberplain}

\def\eg{{\em e.g.,}\xspace}
\def\ie{{\em i.e.,}\xspace}

\usepackage{cleveref}

\crefname{appendix}{Appendix}{Appendices}
\Crefname{appendix}{Appendix}{Appendices}

\begin{document}

\pagestyle{plain}

\maketitle
\thispagestyle{plain}

\begin{abstract}
The growing use of large language models has increased interest in sharing textual data in a privacy-preserving manner. One prominent line of work addresses this challenge through text rewriting under Local Differential Privacy (LDP), where input texts are locally obfuscated before release with formal privacy guarantees. These guarantees are typically expressed by a parameter~$\varepsilon$ that upper bounds the worst-case privacy loss. However, nominal $\varepsilon$ values are often difficult to interpret and compare across mechanisms. In this work, we investigate how to empirically calibrate across text rewriting mechanisms under LDP. We propose \ourmethod, which formulates calibration via a hypothesis-testing framework that instantiates text distinguishability audits in both surface and embedding spaces, enabling empirical assessment of indistinguishability from privatized texts. Applying this calibration to several representative mechanisms, we demonstrate that similar nominal $\varepsilon$ bounds can imply very different levels of distinguishability. Empirical calibration thus provides a more comparable footing for evaluating privacy-utility trade-offs, as well as a practical tool for mechanism comparison and analysis in real-world LDP text rewriting deployments.
\end{abstract}

\section{Introduction} \label{sec:intro}

The widespread adoption of Large Language Models (LLMs) has driven increasingly interactive use of remote services, making the disclosure of personal and potentially sensitive information inevitable and raising persistent privacy concerns. In this user-centric paradigm, the responsibility for data protection, traditionally managed by trusted curators using Differential Privacy (DP)~\citep{dwork2006dp}, shifts to the user. This transition has motivated growing interest in Local Differential Privacy (LDP)~\citep{duchi2013ldp}, where protection is applied directly to user data prior to disclosure. In Natural Language Processing (NLP), this has catalyzed the development of LDP-based text rewriting mechanisms aimed at perturbing user inputs~\citep{meisenbacher2024dpmlm,meisenbacher2025dpst}.

\begin{figure}[t]
\centering
\includegraphics[width=0.98\linewidth]{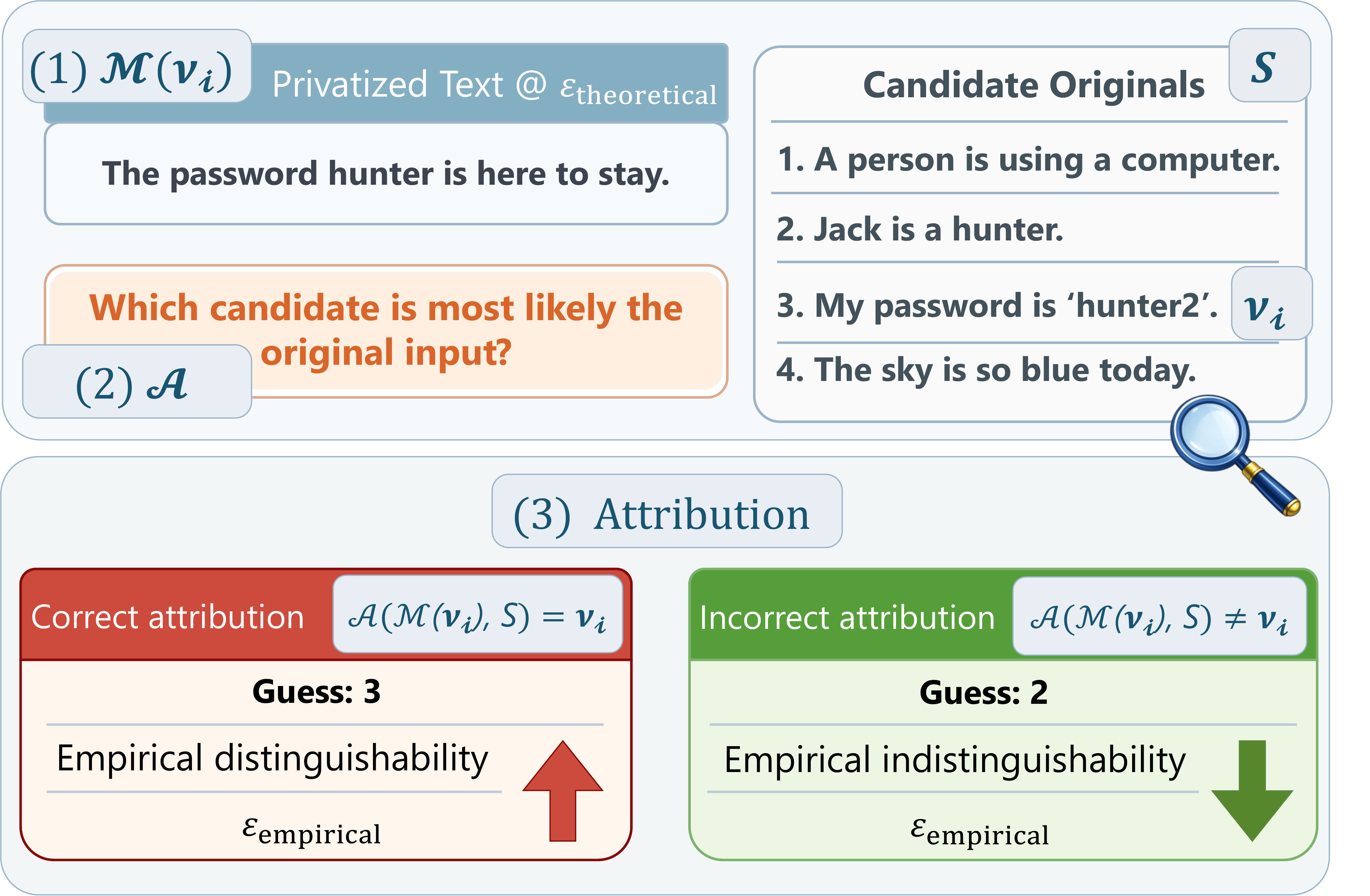}\\
\vspace{-2pt}
\caption{\ourmethodfull for empirical privacy assessment of text rewriting mechanisms. (1)~A mechanism $\mathcal{M}$ produces privatized text at a given privacy budget $\varepsilon_{\text{theoretical}}$; (2)~an adversary $\mathcal{A}$ attempts to identify the true source $v_i$ from a candidate set $S$; (3)~A correct attribution indicates empirical distinguishability, while an incorrect attribution indicates indistinguishability.}
\vspace{-10pt}
\label{fig:teaser}
\end{figure}

\begin{table*}[t]
\renewcommand{\arraystretch}{0.90}
\centering
\small
\begin{tabularx}{\linewidth}{p{2.2cm} X}
\toprule
Method & Rewriting \\
\midrule
\multicolumn{2}{l}{\textcolor{lightblue}{\textit{Ref: I've only used it once and it helped! PLEASE DO NOT SHARE MY NAME OR E-MAIL ADDRESS!}}} \\
\midrule
\multicolumn{2}{c}{\emph{$\varepsilon = 1000$}} \\
\midrule
ADePT     & sues run barcelona because now provided 80 if drive believe flood smart situation sold sign \\
DP-BART   & Click here to view the latest edition of SHARE THIS STORY Helen \\
DP-Paraphrase & I Democratic People helm the White House. foggy skies violence. I dysphagia. I classroom. \\
DP-Prompt & I've only used it once and it helped! Please do not share my name or E-mail address! \\
DP-MLM    & I ve only considered it once and it acted! Please Off Stop Whatsapp Pc * With Help Port! \\
DP-ST     & I have used it. \\
\midrule
\multicolumn{2}{c}{\emph{$\varepsilon = 10$}} \\
\midrule
ADePT     & more growing app \\
DP-BART   & (empty output) \\
DP-Paraphrase & it didinder does.aroo-mute used one person ',it works in an empty store like room a few years while getting on \\
DP-Prompt & Imagesated was 70 contacts too People distinguish maximize\textdegree{} Kunstow shifting valve Elizabeth unlikelyfekt Vienna \\
DP-MLM    & Fascinated enges only surprised it once and it ant! Vacc Ash Hum Appear Executive Morgan 50 Messenger Classification! \\
DP-ST     & I am being used by Asos. \\
\bottomrule
\end{tabularx}
\caption{Example text rewritings at the same theoretical $\varepsilon$ by ADePT~\citep{krishna2021adept}, DP-BART~\citep{igamberdiev2023dpbart}, DP-Paraphrase~\citep{mattern2022dpparaphrase}, DP-Prompt~\citep{utpala2023dpprompt}, DP-MLM~\citep{meisenbacher2024dpmlm} and DP-ST~\citep{meisenbacher2025dpst}.}
\label{tab:trustpilot}
\end{table*}

A central concept in DP is the privacy loss parameter $\varepsilon$, which provides a theoretical upper bound on worst-case distinguishability between outputs generated from any two possible inputs; this bound is commonly reported as a nominal numerical value. Despite its theoretical appeal, interpreting $\varepsilon$ in practice remains challenging.
First, its practical meaning is hard to quantify: knowing that $\varepsilon = 10$ does not directly convey how much information about the original input is leaked in expectation. Second, and more critically, $\varepsilon$ values are not comparable across mechanisms---different text rewriting systems define privacy at different granularities \citep{mattern2022dpparaphrase}, such as the token level~\citep{meisenbacher2024dpmlm} or the sentence level~\citep{igamberdiev2023dpbart}, and even within the same granularity there are differences in tightness of bounds---rendering nominal $\varepsilon$ values incommensurable across methods.\footnote{Cross-granularity conversion via the basic composition theorem~\citep{dwork-roth:2014} is possible but yields loose bounds.} As demonstrated in Table~\ref{tab:trustpilot} (also in~\Cref{tab:trustpilot-full}), mechanisms operating under identical nominal $\varepsilon$ can produce outputs with markedly different levels of observable privacy leakage, underscoring the need for empirical calibration of LDP mechanisms.

In the centralized DP setting, \citet{jayaraman2019evaluating} argued for empirical calibration across DP mechanisms to assess privacy leakage, using membership inference attacks (MIAs) as an evaluation tool~\citep{shokri2017membership, yeom2018privacy}, where an adversary attempts to infer whether a given record was part of the training data. They demonstrated that mechanisms with identical nominal $\varepsilon$ values can exhibit substantially different empirical leakage, indicating that formal guarantees alone are often insufficient for characterizing practical privacy protection or guiding mechanism selection. This observation is linked to a growing body of work on DP auditing~\citep{jagielski2020audit, nasr2021adversary, nasr2023tight, maddock2023canife, steinke2023privacy}, 
which aims to establish tight empirical lower bounds for individual mechanisms using the same kinds of tools (like MIAs) to empirically estimate privacy leakage. A recent approach of auditing LDP protocols was proposed by~\citet{arcolezi2024revealing}, which formalizes output-level auditing through the notion of distinguishability attacks. In this, they aim to assess the distinguishability of inputs directly, positing an adversary trying to predict the user’s input based on the obfuscated output; this aligns with distinguishability as a central concept in DP~\citep{dwork-roth:2014, igamberdiev2023dpbart, meisenbacher2025dpst}. While primarily an auditing framework, their cross-protocol analysis also reveals calibration-relevant insights: mechanisms with identical nominal $\varepsilon$ can exhibit substantially different empirical privacy loss.

However, the framework proposed by \citet{arcolezi2024revealing} was only applied in the context of LDP frequency estimation protocols over tabular data, where outputs lie in low-dimensional, well-defined domains that enable statistical estimation of privacy leakage. Extending this approach to LDP text rewriting is non-trivial and introduces three key challenges: (1) text rewriting mechanisms operate on high-dimensional, discrete inputs whose semantic structure makes re-identification of the source input difficult; (2) candidate selection critically affects estimated leakage in text, unlike tabular settings where random sampling suffices; and (3) standard estimation procedures require a high volume of Monte Carlo trials, typically $10^4$ to $10^6$, which is prohibitively expensive for LDP text rewriting mechanisms that involve running large pretrained language models.

To address these challenges, \updates{we propose \ourmethodfull, \ourmethod}, an empirical calibration framework for LDP text rewriting mechanisms that extends distinguishability-based auditing~\citep{arcolezi2024revealing} to the setting of natural language. We summarize our key contributions and findings below:
\begin{itemize}
    \item We design distinguishability attacks at both surface and embedding levels to assess empirical distinguishability in text rewriting under LDP, addressing the high-dimensional discrete nature of natural language (\Cref{subsec:distinguishability}).
    \item We develop a unified candidate selection strategy that simulates adversaries with varying auxiliary information (\Cref{subsec:worstcase}), and an efficient aggregated estimator that makes evaluation of resource-intensive text rewriting mechanisms practical at scale (\Cref{subsec:aggregation}).
    \item We conduct empirical study that calibrates six representative LDP text rewriting mechanisms and demonstrate that identical nominal $\varepsilon$ can imply substantially different empirical privacy loss, and provide a useful tool that can aid in analyses such as privacy-utility trade-offs, the effect of post-processing, and mechanism-specific auditing (\Cref{sec:experiment}).
\end{itemize}

\section{Related Work} \label{sec:literature}

\noindent \textbf{Differential Privacy.}
DP~\cite{dwork2006dp} provides a rigorous mathematical framework for quantifying privacy leakage in data release and analysis. In the central model, a trusted curator applies randomization to aggregated data before release, with privacy guarantees parameterized by a privacy budget $\varepsilon$. A major application area is privacy-preserving machine learning, where DP-SGD~\cite{abadi2016dpsgd} enables training models on sensitive data by injecting calibrated noise into gradient computations. This has spurred substantial research on improving privacy-utility tradeoffs through refined composition theorems~\cite{mironov2017renyi,pmlr-v80-balle18a}, empirical privacy auditing and calibration methods~\cite{jagielski2020audit,nasr2023tight,steinke2023privacy,pmlr-v235-kaissis24a}. A key insight from this line of work is that nominal $\varepsilon$ values may not accurately reflect empirical privacy loss in practice.

\vspace{5pt}

\noindent \textbf{Local Differential Privacy and Text Rewriting.}
LDP~\cite{duchi2013ldp} extends DP to the decentralized setting, where individuals perturb their own data before transmission, eliminating the need for a trusted curator. LDP mechanisms have been developed for text data~\cite{8970912, hu2024differentially, vu-etal-2024-granularity}, operating at different granularities: token-level methods~\cite{mattern2022dpparaphrase,utpala2023dpprompt,meisenbacher2024dpmlm} apply word-level perturbations, while sentence-level approaches~\cite{krishna2021adept,igamberdiev2023dpbart} generate paraphrases conditioned on semantic representations. However, this diversity in reporting granularity makes direct mechanism comparison and selection non-trivial, as mechanisms with identical nominal $\varepsilon$ may exhibit substantially different empirical privacy-utility tradeoffs.

\vspace{5pt}

\noindent \textbf{Auditing Local Differential Privacy.}
While empirical auditing has been extensively studied for central DP, analogous work for LDP remains sparse and largely confined to tabular domains~\cite{gursoy-2022-10.1109/TIFS.2022.3170242,arcolezi-2023-10.14778/3579075.3579086,arcolezi2024revealing}. \citet{arcolezi2024revealing} proposed a distinguishability-based auditing framework for LDP frequency estimation protocols, where an adversary observes a privatized output $y \sim \mathcal{M}(v_1)$ for a sampled pair $(v_1, v_2) \in \mathcal{V} \times \mathcal{V}$, constructs a mechanism-specific \emph{support set} $\mathcal{S}(y) \subseteq \mathcal{V}$ of values most likely to have generated $y$, and predicts by sampling from $\mathcal{S}(y)$. Empirical distinguishability is estimated as the frequency of correct attribution over many trials.

This framework applies naturally to tabular LDP because domains are finite and enumerable, support sets can be constructed explicitly, and distinguishability is relatively stable across candidate pairs. Text domains, however, present fundamentally different challenges: the input space is combinatorially vast, making exhaustive enumeration intractable; and distinguishability becomes highly heterogeneous, varying dramatically with semantic similarity between candidates. These challenges motivate a new framework for empirical calibration of LDP text rewriting mechanisms.

\section{Methodology}
\label{sec:method}
\vspace{-5pt}
We present \updates{\ourmethodfullbold~(\ourmethod),} for empirical calibration of LDP text rewriting via distinguishability-based assessment.

\subsection{Distinguishability-Based Calibration}
\label{subsec:distinguishability}

Local differential privacy (LDP) aims to ensure that a privatized output does not allow an adversary to reliably distinguish the true input from \textit{any other} neighboring input~\citep{igamberdiev2023dpbart,meisenbacher2025dpst}.\footnote{This differs from metric DP, where level of distinguishability is proportional to distance under some metric.} Formally, an $(\varepsilon, \delta)$-LDP mechanism on $\mathcal M$ guarantees that for any two input texts $x$ and $y$, and any output $z$,
\begin{equation}
\Pr[\mathcal M(x) = z] \le e^{\varepsilon} \cdot \Pr[\mathcal M(y) = z] + \delta,
\end{equation}
where $\delta \ge 0$ is a small failure probability. When $\delta = 0$, this reduces to pure $\varepsilon$-LDP. This definition naturally motivates a distinguishability-based view of privacy calibration. The nominal guarantee may be overly conservative and not reflect the actual privacy loss in practice: a mechanism may report a high $\varepsilon$ while remaining empirically indistinguishable, meaning the true privacy loss is much lower than the nominal bound suggests. Moreover, the tightness of these upper bound guarantees can vary across mechanisms according to how the guarantee is determined (\eg which composition theorem is applied)~\citep{dwork-roth:2014}. 
Since the LDP guarantee is fundamentally defined in terms of distinguishability between inputs, empirical calibration via an adversary's ability to attribute outputs to their source provides a natural and direct means of assessing privacy leakage in a comparable way across mechanisms.

\vspace{5pt}

\noindent \textbf{From tabular to text LDP.} As discussed in \Cref{sec:literature}, existing LDP auditing approaches focus on tabular frequency estimation, where mechanisms operate over discrete categorical values and support sets can be constructed through explicit enumeration.
Text rewriting mechanisms, however, operate in a fundamentally different regime.
Modern text LDP approaches either inject noise into continuous embedding spaces before decoding~\citep{krishna2021adept, igamberdiev2023dpbart} or apply controlled token- or sentence-level perturbations~\citep{mattern2022dpparaphrase, utpala2023dpprompt, meisenbacher2024dpmlm, meisenbacher2025dpst}.
The resulting output space is large, structured, and semantically meaningful, making exhaustive enumeration or exact input reconstruction infeasible.

\vspace{5pt}

\noindent \textbf{Candidate-based distinguishability formulation.}
To obtain a practical and mechanism-agnostic calibration framework, we adopt a candidate-based formulation. Let $\mathcal{M}$ denote an LDP text rewriting mechanism that produces privatized output $y = \mathcal{M}(v_i)$ from a true source input $v_i$. Given $y$ and a finite candidate set $\mathcal{S} = \{v_1, \dots, v_k\}$ with $v_i \in \mathcal{S}$, the goal is to infer which candidate most plausibly generated $y$.
This formulation avoids explicit enumeration of the full output space while remaining faithful to the indistinguishability requirement of LDP.\footnote{This idea has been applied in tasks where even human ability to reconstruct is a challenge because of the candidate set size, such as in assessing privacy of reconstructed images \citep{sun2023privacy} or in unsupervised summarisation \citep[\S4.5]{zhuang-etal-2022-learning}.  Evaluating reconstructions is also challenging because existing metrics do not necessarily capture actual privacy leakage \citep{sun2023privacy,faustini-etal:2025}.} The threat model assumes an adversary that has some auxiliary information \citep[\S2.2]{dwork-roth:2014} regarding inputs, in the form of candidate sets that are constructed by combining $v_i$ with additional samples drawn from the same domain. In this work we draw the additional samples (distractors) from a training dataset, although they could come from elsewhere. The role of candidate construction and its implications for privacy calibration are discussed in~\Cref{subsec:worstcase}.

\vspace{5pt}

\noindent \textbf{Distinguishability attacks for text rewriting.}
We instantiate two distinguishability attacks that exploit the semantic structure of text and apply across a wide range of LDP rewriting mechanisms.

\textbf{Text-based semantic attribution.}
Text rewriting mechanisms operate on discrete linguistic representations, for which no canonical distance or similarity function captures semantic relatedness. From the adversary's perspective, attribution naturally proceeds by comparing surface-level cues such as wording, phrasing, and syntactic structure. We operationalize this by instantiating $f(\cdot,\cdot)$ using a large language model, which approximates such judgments at scale. Given a candidate set $\mathcal{S}$ and an observed output $y$, the predicted source is
\begin{equation}
\mathcal{A}_{\mathrm{text}}(y) = \arg\max_{v \in \mathcal{S}} f(y, v).
\end{equation}
The attack succeeds if the rewritten text remains more semantically aligned with its true source $v_i$ than with alternative candidates.

\textbf{Embedding-based attribution.}
Pretrained embedding models encode textual semantics into continuous vector representations by design, inducing a geometry over the input space that enables attribution via standard distance measures. Moreover, the post-processing property of DP guarantees that any computation applied to a privatized output remains differentially private~\citep{dwork-roth:2014}, formally grounding embedding-space auditing. Let $\phi(\cdot)$ denote a fixed embedding function and $d(\cdot,\cdot)$ a distance metric. The predicted source is
\begin{equation}
\mathcal{A}_{\mathrm{emb}}(y) = \arg\min_{v \in \mathcal{S}} d\!\left(\phi(y), \phi(v)\right).
\end{equation}
These two attacks are complementary: text-based attribution captures surface-level leakage as a human adversary would perceive it, while embedding-based attribution probes deeper semantic structure that may persist even when surface form is substantially altered.
These give two perspectives on calibration; we would expect them to align in calibrating mechanisms.


\subsection{Candidate Selection for Calibration}
\label{subsec:worstcase}

A defining characteristic of LDP is that every element of the input domain is considered adjacent to every other, meaning the mechanism must make any two inputs indistinguishable~\citep{igamberdiev2023dpbart,meisenbacher2025dpst}. While this is largely inconsequential for calibrating LDP mechanisms over discrete, unstructured domains such as tabular data~\citep{arcolezi2024revealing}, it has critical implications for text rewriting, where the input space is structured and semantically rich. In such settings, different choices of neighboring inputs could induce substantially different levels of distinguishability, which might in turn affect empirical privacy calibration.  Below we propose a framework considering a range of candidate selection strategies to demonstrate that our empirical privacy calibration is relatively robust to the choice.

\subsubsection{Why Candidate Selection Matters in Text LDP}
\label{subsec:why-worstcase}

In the classical privacy problem of designing private frequency estimation protocols on tabular data, typical mechanisms like (generalized) randomized response will select outputs other than the true one uniformly at random. Any choice of distractor for the distinguishability adversary in prior LDP auditing approaches is therefore satisfactory, since any pair of (true, distractor) values yields comparable outcomes. This symmetry underlies the observation that, unlike central DP auditing, LDP auditing for frequency estimation does not depend on identifying specific worst-case inputs~\citep{arcolezi2024revealing}.

Text domains, by contrast, exhibit rich semantic structure. Input texts may differ substantially in lexical overlap, semantic similarity, or embedding distance---differences that can persist after privatization, causing empirical distinguishability to vary with the choice of candidates. Ignoring candidate selection may thus yield overly optimistic estimates that underestimate the distinguishability a realistic observer could achieve. Candidate selection should therefore be treated as an integral component of text LDP calibration, consistent with the principle that stronger attacks reveal more about practical privacy loss~\citep{jagielski2020audit}.

\subsubsection{Candidate Selection Strategies}
\label{subsec:sampling}
We consider three intuitive configurations: (i) \textit{uniform sampling}, which selects candidates without regard to semantic proximity; (ii) \textit{more-similar sampling}, which draws candidates close to the true source, imposing a harder distinguishability task on the adversary; and (iii) \textit{more-diverse sampling}, which draws semantically dissimilar candidates, making attribution easier for the adversary and thus surfacing greater distinguishability pressure for a more informative calibration estimate.

We propose \textit{Probabilistic Transition Sampling} (detailed in~\Cref{appendix:sampling}), a unified strategy that interpolates between these configurations via a temperature parameter $\lambda$: $\lambda < 0$ gives more-diverse sampling, $\lambda > 0$ more-similar sampling, and $\lambda \to 0$ uniform sampling~\citep{arcolezi2024revealing}.

We adopt more-diverse sampling ($\lambda < 0$) as our primary configuration, as it maximizes distinguishability pressure to the adversary. The overall pattern of calibration is consistent across configurations (verified empirically in~\Cref{sec:ablation}), suggesting our calibration conclusions are robust to the specific choice of candidate selection strategy.


\subsection{Efficient Estimation}
\label{subsec:aggregation}

Prior distinguishability-based auditing estimates empirical privacy loss by tracking two types of events: a true positive (TP), where the adversary correctly identifies the true source from its privatized output, and a false positive (FP), where the adversary incorrectly attributes a different input. This yields symmetric hypothesis testing procedures that typically require $10^4$ to $10^6$ Monte Carlo trials to achieve reliable estimates~\citep{jagielski2020audit,arcolezi2024revealing}. While this cost is manageable for tabular mechanisms operating on numerical data, text rewriting mechanisms require loading and querying large pretrained models at each trial, making the full symmetric procedure prohibitively expensive in practice.

We observe that this symmetric formulation offers an opportunity for efficiency gains: a false positive under one hypothesis corresponds exactly to a true positive under the swapped hypothesis, meaning all relevant distinguishability information can be captured through aggregated success events alone, without separate TP and FP estimation. We derive a reduced-form estimator that depends only on aggregated success events, preserving estimation correctness while halving the number of required model queries. Following~\citet{jagielski2020audit} and \citet{arcolezi2024revealing}, we rely on Monte Carlo sampling to estimate success probabilities and adopt the Clopper-Pearson method~\citep{clopper1934use} to construct exact confidence intervals. A formal justification of this reduction and the resulting estimator is provided in~\Cref{app:proof}; the complete estimation procedure is given in Algorithm~\ref{alg:audit} (\Cref{app:algo}).

\section{Experimental Setup and Results}
\label{sec:experiment}
\vspace{-5pt}
\subsection{Experimental Setup} \label{sec:setup}

\paragraph{Datasets.} We evaluate on four datasets commonly used in LDP text rewriting research: ATIS~\citep{dahl1994atis}, SST-2~\citep{socher2013sst2}, SNIPS~\citep{coucke2018snips}, and Trustpilot~\citep{hovy2015trustpilot}. These datasets span different text domains and sentence lengths, providing a diverse testbed for calibration across mechanisms.

\vspace{4pt}

\noindent \textbf{LDP Text Rewriting Mechanisms.} We evaluate six representative mechanisms spanning two main technical paradigms. ADePT and DP-BART privatize text by injecting Laplace or Gaussian noise directly into continuous embedding representations, while the remaining four methods realize privacy via the Exponential Mechanism (EM): DP-Paraphrase, DP-Prompt, and DP-MLM control the temperature sampling of token generation, and DP-ST replaces semantic triples with candidates drawn from an auxiliary corpus.
\begin{enumerate}
\item \textbf{ADePT}~\citep{krishna2021adept,habernal-2021-differential}: an RNN-based autoencoder that privatizes text by injecting calibrated noise into the encoder's latent representation before decoding. 
\item \textbf{DP-Paraphrase}~\citep{mattern2022dpparaphrase}: applies temperature sampling during autoregressive generation in a GPT-2~\citep{Radford2019LanguageMA} to provide token-level guarantees.
\item \textbf{DP-BART}~\citep{igamberdiev2023dpbart}: adds noise to the sentence-level latent embedding of a BART autoencoder~\citep{lewis2020bart}. We use the DP-BART-PR+ variant.
\item \textbf{DP-Prompt}~\citep{utpala2023dpprompt}: achieves token-level DP guarantees during text generation by treating temperature sampling as an exponential mechanism, using FLAN-T5~\citep{chung2024scaling} as the underlying model.
\item \textbf{DP-MLM}~\citep{meisenbacher2024dpmlm}: performs token-level privatization via the exponential mechanism over candidate tokens generated by a masked language model, using RoBERTa~\citep{liu2019roberta} as the backbone.
\item \textbf{DP-ST}~\citep{meisenbacher2025dpst}: extracts semantic triples from the input, privatizes each via the exponential mechanism over an auxiliary corpus, and reconstructs coherent output via LLM paraphrasing.
\end{enumerate}

\begin{figure*}[t]
\centering
\includegraphics[width=0.9\linewidth]{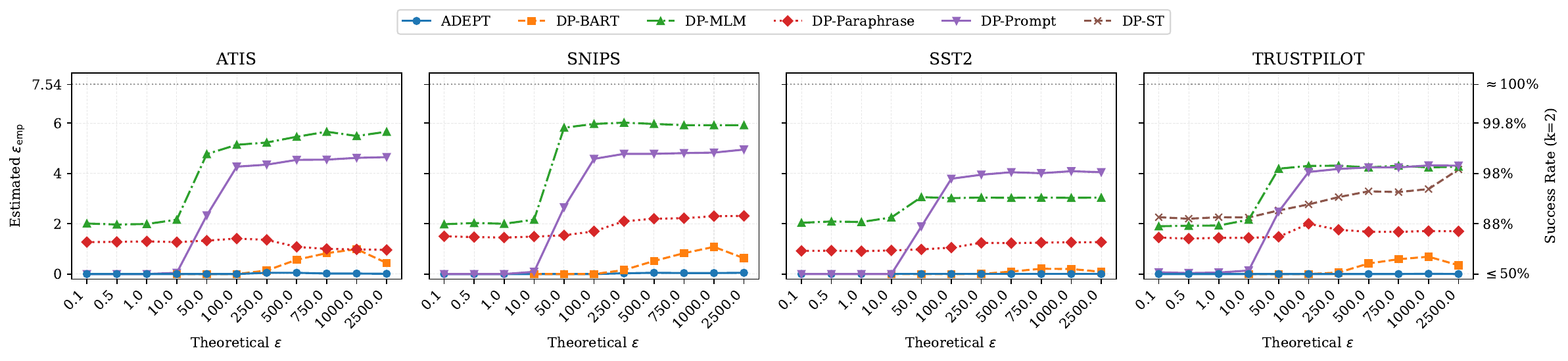}
\vspace{-10pt}
\caption{
    Empirical calibration results under the LLM distinguishability attack across datasets. The x-axis shows nominal $\varepsilon$ and the y-axis shows estimated empirical privacy loss $\varepsilon_{\mathrm{emp}}$. The dashed horizontal line marks the finite-sample ceiling (${\approx}7.54$, $k=2$, $T=10^4$).
}
\label{fig:llm-results}
\end{figure*}
\vspace{-1pt}

\noindent \textbf{Distinguishability Attacks.} We consider two attacks as introduced in~\Cref{subsec:distinguishability}. For text-based attribution, we use Qwen3-8B~\citep{yang2025qwen3} as the LLM judge, prompting it with the privatized output and $k$ candidate inputs and asking it to identify the most plausible source; the full prompt is provided in~\Cref{app:prompt}. For embedding-based attribution, we encode all texts using \texttt{all-mpnet-base-v2}~\citep{song2020mpnet} and identify the source candidate via cosine distance in embedding space. For ADePT and DP-BART, which operate explicitly in embedding space, we additionally evaluate distinguishability on internal embeddings, directly comparing pre- and post-noise latent representations. We do this as part of the validation of our approach, as it allows us to assess the most direct connection between the mechanism and distinguishability: higher noise (lower $\varepsilon$) should correspond monotonically to worse distinguishability, with no post-processing (\eg decoding) to interfere. We refer to these three as the \textbf{LLM}, \textbf{external embedding}, and \textbf{internal embedding} distinguishability attacks throughout the remainder of this paper.

\noindent \textbf{Implementation.} For ADePT and DP-BART, we use the implementation and checkpoints from~\citet{igamberdiev2023dpbart}; for DP-Paraphrase, DP-Prompt, and DP-MLM, the code from~\citet{meisenbacher2024dpmlm}; and for DP-ST, the code from~\citet{meisenbacher2025dpst}. In all cases, we retain the default hyperparameters from the respective works. We evaluate with $T = 10^4$ and $T = 10^6$ Monte Carlo trials using a distinct random seed per trial and find consistent results; we therefore report at $T = 10^4$ to keep overhead tractable. Our pipeline extends to any candidate set size $k$; we report at $k = 2$ for comparability with~\citet{arcolezi2024revealing} and verify consistent trends at $k \in \{4, 8, 16\}$ (see Appendix~\ref{app:implementation}).

\vspace{4pt}

\noindent \textbf{Downstream Utility and Privacy Evaluation.}
To assess the privacy-utility tradeoff across mechanisms, we follow the evaluation protocol of~\citet{meisenbacher2025dpst}. For each mechanism, we fine-tune a DeBERTa-v3-base~\citep{he2021deberta} classifier for one epoch on the privatized training set and evaluate on the original clean test set. On Trustpilot, we evaluate both a utility task (binary sentiment classification) and a privacy task (binary gender attribute inference), where lower F1 on the latter indicates stronger resistance to attribute inference. On SNIPS, we evaluate the utility task (multi-class intent classification). All experiments are repeated over 3 random seeds and we report mean F1 with standard deviation.

\subsection{Results} \label{sec:main_results}

\vspace{-5pt}

\noindent \textbf{Interpreting the Calibration Scale.}
Figure~\ref{fig:llm-results} presents the empirical calibration results. We clarify how to interpret $\varepsilon_{\mathrm{emp}}$ and the secondary axis. By~\eqref{eq:binary-estimator}, $\varepsilon_{\mathrm{emp}}$ depends directly on $\hat{p}_0$, the Clopper-Pearson confidence lower bound on the true success probability: when the success rate is at or below 50\%, the adversary performs no better than random guessing, $\hat{p}_0$ is even lower, and $\varepsilon_{\mathrm{emp}} = 0$, indicating strong empirical privacy protection. At the other extreme, even when the success rate reaches 100\%, $\hat{p}_0$ remains strictly below 1 due to the Clopper-Pearson confidence interval with finite samples, yielding a finite ceiling on $\varepsilon_{\mathrm{emp}}$: with $k=2$ and $T=10{,}000$ trials, this ceiling is approximately 7.54, represented by the dashed horizontal line. A mechanism reaching this ceiling is empirically fully distinguishable. This bounded scale provides a common empirical reference for comparing mechanisms across different nominal $\varepsilon$ values and datasets.

\begin{figure*}[t]
\centering
\includegraphics[width=0.87\linewidth]{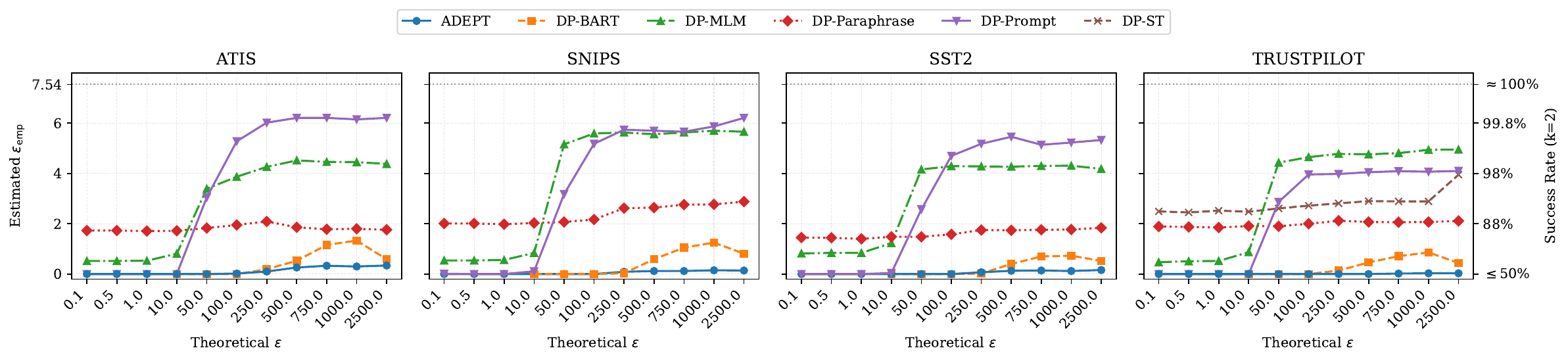}\\
\includegraphics[width=0.87\linewidth]{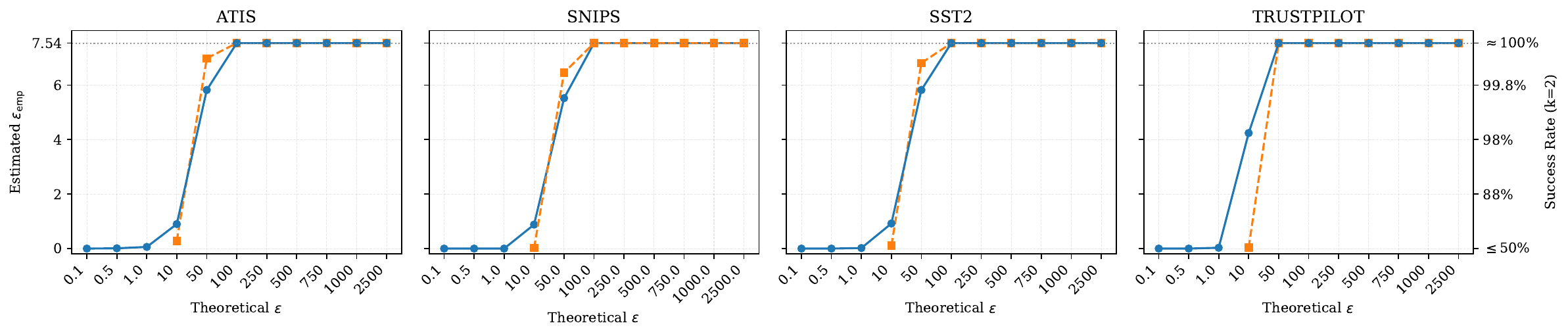}
\vspace{-10pt}
\caption{
    Empirical calibration results under the \textbf{external} (\textit{top}) and \textbf{internal} 
    (\textit{bottom}) embedding distinguishability attacks across datasets. Axes and ceiling as in Figure~\ref{fig:llm-results}.
}
\label{fig:embedding-results}
\end{figure*}

\vspace{3pt}

\noindent \textbf{Text-Space Calibration Results.}
Figure~\ref{fig:llm-results} presents empirical calibration results under the LLM attack across all four datasets. 
Broadly, the ranking aligns with impressions from examples in Tables~\ref{tab:trustpilot} and \ref{tab:trustpilot-full}: ADePT and DP-BART provide most protection, DP-Paraphrase somewhat less, and DP-Prompt less again. We observe a clear separation between two groups of mechanisms: ADePT and DP-BART consistently yield $\varepsilon_{\mathrm{emp}}$ values close to zero, while DP-MLM, DP-Paraphrase, DP-Prompt, and DP-ST exhibit substantially higher $\varepsilon_{\mathrm{emp}}$ that rises with $\varepsilon$. This reflects a fundamental difference in privacy budget reporting granularity across mechanisms. ADePT and DP-BART inject noise into sentence embeddings and report $\varepsilon$ at the sentence level, whereas DP-Paraphrase, DP-Prompt, and DP-MLM are token-level mechanisms that report $\varepsilon$ per token: by the basic composition~\citep{dwork-roth:2014}, rewriting a sentence of $n$ tokens consumes a total budget of $n\varepsilon$~\citep{meisenbacher2024dpmlm}, resulting in a substantially looser sentence-level guarantee.

Even among token-level methods, the effective sentence-level budget varies significantly, as different tokenization strategies yield different numbers of sub-tokens and hence different total budgets (see~\Cref{tab:tokenization-comparison}). DP-ST operates at the level of semantic triples---a granularity between tokens and sentences---and its $\varepsilon_{\mathrm{emp}}$ values fall correspondingly between the two groups.

This heterogeneity means that identical nominal $\varepsilon$ values imply very different levels of actual privacy protection, and direct numerical comparison across mechanisms requires careful consideration of reporting granularity. Although cross-level conversion via the basic composition theorem is possible, the resulting bounds are known to be loose, and as noted by~\citet{mattern2022dpparaphrase}, there are practical reasons to retain token-level reporting, such as lower computational cost and independence from the target dataset. This incommensurability across reporting granularities is precisely where empirical calibration provides the most value: by directly measuring distinguishability under a common adversarial framework, our approach enables meaningful comparison across mechanisms regardless of how their nominal $\varepsilon$ is defined. Figure~\ref{fig:cross-convention} in~\Cref{app:add-results} presents a cross-convention comparison on the ATIS dataset, converting token-level mechanisms to sentence-level equivalents via the basic composition theorem using mean sentence token length, and the trend and ranking remain consistent with the main results.

We further note two anomalies in the calibration curves. DP-BART exhibits a non-monotonic decrease in $\varepsilon_{\mathrm{emp}}$ as $\varepsilon$ increases from 1000 to 2500, and DP-Paraphrase shows an unexpectedly flat response across all nominal $\varepsilon$ values. We investigate both behaviors in~\Cref{sec:ablation}.

\begin{figure*}[t]
\centering
\includegraphics[width=0.85\linewidth]{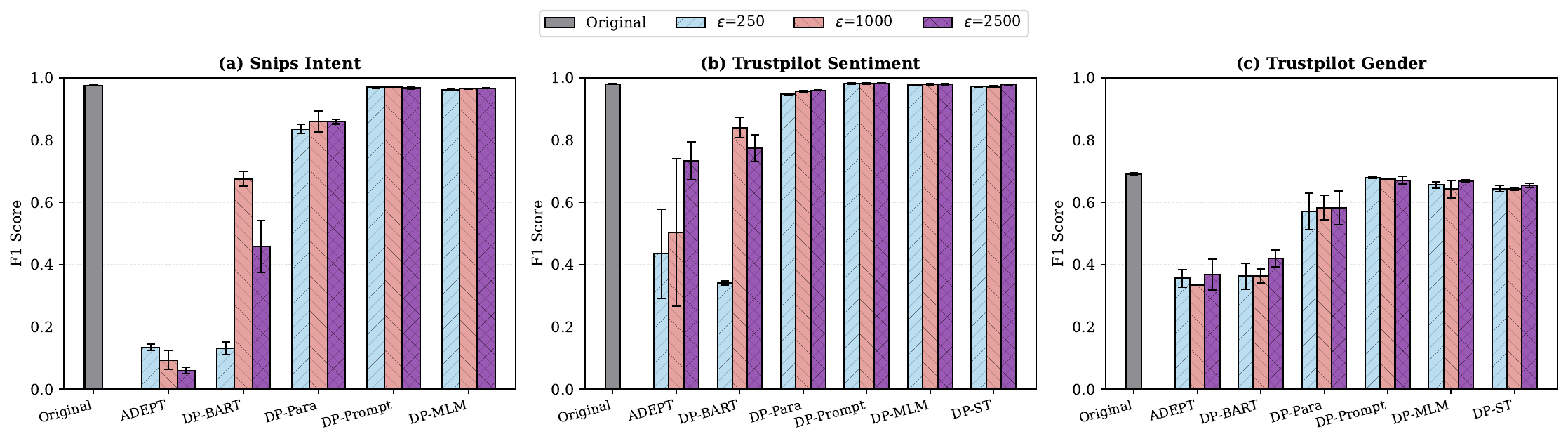}
\vspace{-10pt}
\caption{
    Downstream utility and privacy attribute protection across LDP text rewriting methods, evaluated at $\varepsilon \in \{250, 1000, 2500\}$ with error bars over 3 random seeds. (a) SNIPS intent classification and (b) Trustpilot sentiment report utility F1 (higher is better). (c) Trustpilot gender attribute inference F1 measures resistance to private attribute inference (lower is better).
}
\label{fig:downstream-utility}
\end{figure*}

\begin{table}[h]
\centering
\small
\begin{tabularx}{\linewidth}{lX}
\toprule
& \textbf{Text} \\
\midrule
\multicolumn{2}{l}{\textit{Case 1: LLM correct, embedding wrong (DP-MLM)}} \\
\midrule
Rewritten & moving from india to salt washing \\
\hline
True input & flights from indianapolis to seattle washington \\
Distractor & what is bna \\
\midrule
\multicolumn{2}{l}{\textit{Case 2: LLM wrong, embedding correct (DP-Prompt)}} \\
\midrule
Rewritten & the following restrictions are not included in the text \\
\hline
True input & explain restriction ap 57 \\
Distractor & what is the latest flight from boston to denver that serves a meal \\
\bottomrule
\end{tabularx}
\vspace{-6pt}
\caption{Complementary behavior of LLM and embedding attacks on ATIS.}
\vspace{-12pt}
\label{tab:attack-complementarity}
\end{table}

\vspace{2pt}

\noindent \textbf{Embedding-Space Calibration Results.} Figure~\ref{fig:embedding-results} presents results under the external and internal embedding distinguishability attacks. The external embedding attack yields an mechanism ranking consistent with the LLM attack, with the two-group separation between sentence-level and token-level methods preserved across all datasets. Within the token-level group, we observe a subtle complementarity between the two attack types: on ATIS, DP-MLM ranks higher than DP-Prompt under the LLM attack but lower under the external embedding attack. As illustrated in Table~\ref{tab:attack-complementarity}, this reflects a difference in what each attack captures---DP-MLM's token substitutions preserve surface structure while shifting semantics, making outputs more identifiable by surface-sensitive LLM judgments but less so in embedding space; DP-Prompt produces greater surface variation while retaining key semantic overlapping, which embedding-based attribution exploits more effectively. For ADePT and DP-BART, the perturbed embedding is the direct mechanism output, with decoding serving as post-processing. We additionally evaluate an internal embedding attack directly on the pre- and post-noise latent representations. Most importantly, as shown in the bottom panel of Figure~\ref{fig:embedding-results}, the behavior is monotonic with respect to $\varepsilon$: without interference by post-processing, distinguishability attacks act as expected, being more successful as less noise is added.  With our data, both mechanisms reach near-ceiling distinguishability at the latent level. The fact that DP-BART continues to be more distinguishable under internal embeddings for small quantities of noise ($\varepsilon = 2500$) while this is not the case under LLM or external embeddings means that we can say something about the privacy effect of the post-processing decoding; the standard DP guarantee regarding post-processing only guarantees that arbitrary computations on DP outputs remain DP, without providing any quantification of effects. 

\vspace{3pt}

\noindent \textbf{Privacy-Utility Tradeoff Signals.} Figure~\ref{fig:downstream-utility} presents downstream utility and private attribute inference results across mechanisms. These results demonstrate that $\varepsilon_{\mathrm{emp}}$ provides a more informative signal of practical privacy-utility tradeoffs than nominal $\varepsilon$. For ADePT and DP-BART, their near-zero $\varepsilon_{\mathrm{emp}}$---despite operating at nominal $\varepsilon$ as large as 1000 or 2500---better reflects their substantial impact on downstream utility. Conversely, the higher $\varepsilon_{\mathrm{emp}}$ of token-level mechanisms, despite their relatively smaller nominal $\varepsilon$, better captures their stronger utility preservation alongside weaker attribute privacy protection.

\vspace{-5pt}

\subsection{Ablation Studies} \label{sec:ablation}

\vspace{-2pt}

\begin{figure}[h]
\centering
\includegraphics[width=0.8\linewidth]{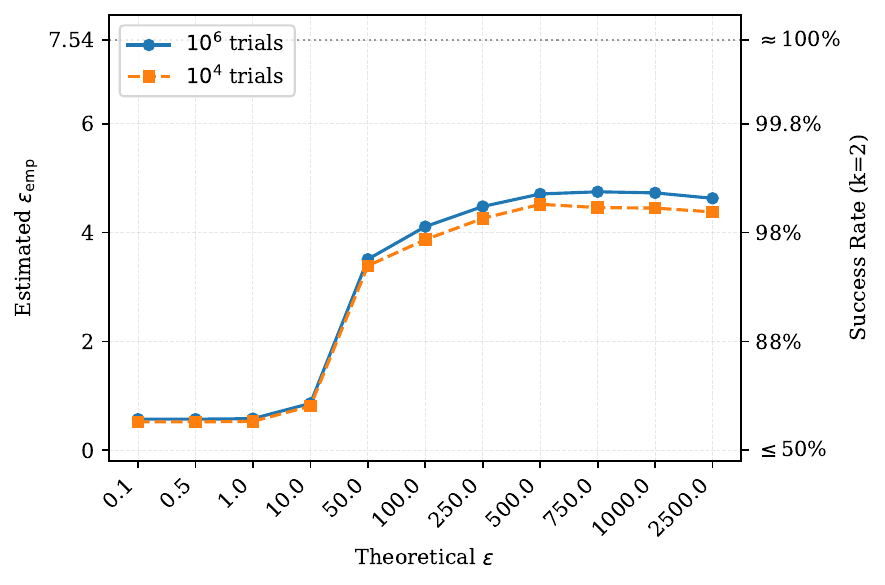}
\vspace{-10pt}
\caption{Empirical calibration results under $T=10^4$ and $10^6$ Monte Carlo trials with DP-MLM on ATIS, showing consistent trends across $\varepsilon$ values.}
\label{fig:monte-carlo-comparison}
\vspace{-2pt}
\end{figure}

\noindent \textbf{Monte Carlo Trial Count.}
Figure~\ref{fig:monte-carlo-comparison} compares calibration results under $T=10^4$ and $T=10^6$ Monte Carlo trials using DP-MLM as a representative method, showing consistent trends across both settings. We thus report all results at $T=10^4$ to keep computational overhead tractable. For DP-MLM, $T=10^6$ requires approximately 18 hours on a single NVIDIA V100 32GB GPU, whereas $T=10^4$ completes within 10 minutes for the embedding attack and 30 minutes for the LLM attack, with all mechanisms finishing within one hour. This reduction makes exhaustive evaluation across six mechanisms, four datasets, and multiple distinguishability attacks practically feasible without sacrificing statistical reliability.
\vspace{-5pt}

\begin{figure*}[t]
    \centering
    \includegraphics[width=0.86\linewidth]{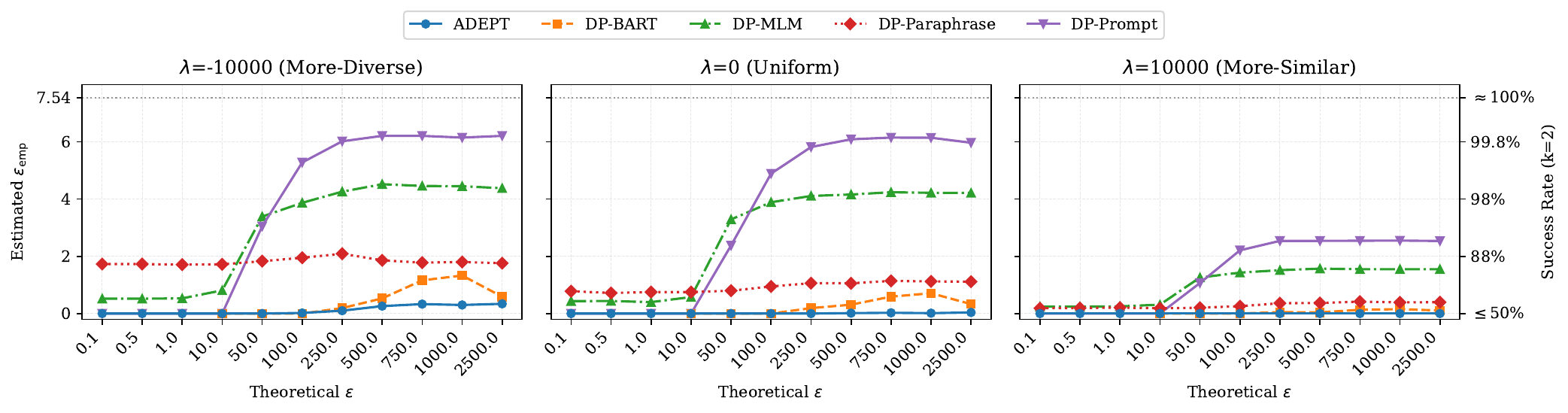}
    \vspace{-10pt}
    \caption{Results under three candidate selection strategies on the ATIS dataset: (a) more-diverse sampling ($\lambda$=-10,000), (b) uniform random sampling ($\lambda$=0), and (c) more-similar sampling ($\lambda$=10,000). While the empirical privacy loss varies depending on the auxiliary information available to the adversary, the relative ranking and trends across different LDP mechanisms remain consistent.}
    \label{fig:ablation-lambda}
\end{figure*}

\vspace{5pt}

\noindent \textbf{Effect of Candidate Set Size.}
Figure~\ref{fig:ablation-study} (top) reports $\varepsilon_{\mathrm{emp}}$ under varying candidate set sizes $k \in \{2, 4, 8, 16\}$ on ATIS, showing that while larger $k$ raises the ceiling of $\varepsilon_{\mathrm{emp}}$ (as the prior odds increase with more candidates), the overall trend remain stable. This suggests that the choice of $k$ affects the scale of the estimate but not the qualitative conclusions drawn from the calibration.

Figure~\ref{fig:ablation-study} (bottom) further validates this stability by reporting results under $k=4$ across all four datasets, under the external embedding attack. The mechanism ranking observed under $k=2$ is preserved, indicating that our calibration findings are robust to the amount of auxiliary information available to the adversary.

\vspace{5pt}

\noindent \textbf{Effect of Sampling Temperature.}
Figure~\ref{fig:ablation-lambda} compares three sampling configurations: more-similar sampling ($\lambda = 10,000$), uniform random sampling ($\lambda = 0$), and more-diverse sampling ($\lambda = -10,000$), representing adversaries with varying levels of prior knowledge. While the specific values of $\varepsilon_{\mathrm{emp}}$ differ across configurations, the overall ranking remains consistent, suggesting that our calibration conclusions are not sensitive to the choice of candidate selection strategy.

\vspace{5pt}

\noindent \textbf{Effect of Efficient Estimation.}
We examine the effect of our efficiency improvement described in Section~\ref{subsec:aggregation}, which avoids the explicit calculation of FPs required by~\citet{arcolezi2024revealing}. Figure~\ref{fig:baseline-comparison} compares three calibration configurations on ATIS under the external embedding model: the baseline of~\citet{arcolezi2024revealing} with uniform random sampling ($\lambda=0$), the same baseline with more-diverse sampling ($\lambda=-10,000$), and our proposed method with more-diverse sampling. The key point here is that the original and efficient versions both at ($\lambda=-10,000$) are close to identical; any differences are much smaller than those introduced by switching to uniform random sampling, confirming that our efficiency improvement does not meaningfully affect calibration results.

\vspace{5pt}

\noindent \textbf{Mechanism-Specific Behaviors.} In Figure~\ref{fig:llm-results}, we observe two mechanism-specific behaviors that deviate from the expected monotonic trend.

For DP-BART, $\varepsilon_{\mathrm{emp}}$ decreases as $\varepsilon$ increases from 1000 to 2500. Table~\ref{tab:dpbart-variance} shows that at $\varepsilon=2500$, outputs exhibit high variance and frequently diverge into generic content, while at $\varepsilon=1000$ outputs remain semantically consistent. As shown in Figure~\ref{fig:dpbart-snr-analysis}, DP-BART clips embeddings to a fixed L2 norm prior to noise injection; at higher $\varepsilon$, the increased signal-to-noise ratio keeps the perturbed embedding close to the clipped representation, causing the decoder to fall back on memorized content~\citep{carlini2021memorization} rather than faithfully reconstructing the input. The resulting outputs are semantically disconnected from the source and harder to attribute. This is consistent with the utility drop at $\varepsilon=2500$ in Figure~\ref{fig:downstream-utility}.

For DP-Paraphrase, $\varepsilon_{\mathrm{emp}}$ remains flat across all nominal $\varepsilon$ values. As shown in Table~\ref{tab:tokenization-comparison}, outputs across $\varepsilon \in \{250, 1000, 2500\}$ are consistently disconnected from the source but share similar stylistic patterns. We attribute this to the GPT-2 paraphraser fine-tuned on SNLI entailment pairs~\citep{meisenbacher2024dpmlm}: as shown in Figure~\ref{fig:position0_prob}, source-relevant tokens have negligible probability mass while SNLI-style tokens dominate at every temperature setting, rendering outputs insensitive to privacy budget changes.

Overall, then, our distinguishability calibration framework can also be useful in an auditing context, as it facilitates the identification of possible issues with specific privacy mechanisms.

\section{Conclusion} \label{sec:conclusion}
\vspace{-5pt}
We propose \ourmethod, an empirical calibration framework that extends distinguishability-based auditing from tabular LDP to text rewriting. By addressing the challenges of high-dimensional discrete text and candidate selection, and improving efficiency, \ourmethodname enables meaningful comparison across mechanisms with incommensurable nominal privacy guarantees. Our evaluation across six mechanisms reveals that identical nominal $\varepsilon$ values can yield substantially different levels of privacy in practice. These findings demonstrate that empirical calibration is important for better understanding nominal guarantees, and that \ourmethodname offers a practical tool for mechanism comparison and analysis in real-world LDP text rewriting deployments, as well as leading to insights into mechanism behaviors previously not noted.

\clearpage

\bibliography{tacl2021}
\bibliographystyle{acl_natbib}

\clearpage
\appendix

\crefalias{section}{appendix}
\crefalias{subsection}{appendix}
\crefalias{subsubsection}{appendix}

\section{Proof of the Reduced-Form Privacy Loss Estimator}
\label{app:proof}

We provide a formal justification for the reduced-form estimator derived in~\Cref{subsec:aggregation}. The key insight is that a false positive (FP) event under one hypothesis corresponds exactly to a true positive (TP) event under the swapped hypothesis, meaning all relevant distinguishability information can be captured through aggregated success events alone, without separately tracking TP and FP rates. We first establish the notation used throughout this proof.

\paragraph{Data and mechanisms.}
$\mathcal{X} = \{x_1, \dots, x_N\}$ denotes the input dataset. $\mathcal{M}$ denotes the LDP text rewriting mechanism and $\mathcal{A}$ denotes the distinguishability attack. $S = \{s_1, \dots, s_k\}$ denotes a specific candidate set of size $k$, drawn via the probabilistic transition sampling in~\Cref{subsec:sampling}. Let $\mathcal{C}_k = \{S \subseteq \mathcal{X} : |S| = k\}$ denote the collection of all size-$k$ subsets of $\mathcal{X}$. 

\paragraph{Random variables.}
$\mathrm{Cand}$ denotes the candidate set, sampled as explained in Section~\ref{subsec:worstcase}.
$\mathrm{IN}$ denotes the true input, the target drawn uniformly from $S$ in line~11 of Algorithm~\ref{alg:audit}, \ie
\begin{equation}
\mathbb{P}(\mathrm{IN}=v \mid \mathrm{Cand}=S) = \frac{1}{k} \quad \text{for all } v \in S.
\label{eq:uniform}
\end{equation}

$\mathrm{guess} := \mathcal{A}(\mathcal{M}(\mathrm{IN}))$ denotes the output of $\mathcal{A}$ applied to the privatized output $\mathcal{M}(\mathrm{IN})$, taking values in $S$.

\paragraph{Events.}
For any $v, v' \in S$ with $v \neq v'$, a true positive (TP) event is $(\mathrm{Cand}=S \wedge \mathrm{IN}=v \wedge \mathrm{guess}=v)$, \ie the adversary correctly identifies the true source. A false positive (FP) event is $(\mathrm{Cand}=S \wedge \mathrm{IN}=v' \wedge \mathrm{guess}=v)$, \ie the adversary attributes $v$ when the true source is $v'$.

\paragraph{Setup.}
We define the overall success and failure events as unions over all possible candidate sets $S \in \mathcal{C}_k$ and all possible true inputs $v \in S$:
\begin{align*}
A &= \text{success}\\
&= \bigcup_{S \in \mathcal{C}_k} \bigcup_{v \in S} (\mathrm{Cand}=S \wedge \mathrm{IN}=v \wedge \mathrm{guess}=v), \\
\bar{A} &= \text{failure}\\
&= \bigcup_{S \in \mathcal{C}_k} \bigcup_{v \in S} (\mathrm{Cand}=S \wedge \mathrm{IN}=v \wedge \mathrm{guess} \neq v).
\end{align*}

By construction, $A$ and $\bar{A}$ are complementary, \ie $\mathbb{P}(A) + \mathbb{P}(\bar{A}) = 1$.

\paragraph{Step 1: Rewriting the failure event.}
We rewrite $\bar{A}$ in triple-union form. First, we expand $\mathrm{guess} \neq v$ as a union over all distinct $v' \in S \setminus \{v\}$:
\begin{multline}
\bar{A} = \mathop{\bigcup}_{S \in \mathcal{C}_k}\; \mathop{\bigcup}_{v \in S}\; \mathop{\bigcup}_{v' \in S \setminus \{v\}} \\
(\mathrm{Cand}=S \wedge \mathrm{IN}=v \wedge \mathrm{guess}=v').
\end{multline}

Swapping the order of the two inner unions:
\begin{multline}
\bar{A} = \mathop{\bigcup}_{S \in \mathcal{C}_k}\; \mathop{\bigcup}_{v' \in S}\; \mathop{\bigcup}_{v \in S \setminus \{v'\}} \\
(\mathrm{Cand}=S \wedge \mathrm{IN}=v \wedge \mathrm{guess}=v').
\end{multline}

Relabeling $v \leftrightarrow v'$ under the total union yields:
\begin{multline}
\bar{A} = \mathop{\bigcup}_{S \in \mathcal{C}_k}\; \mathop{\bigcup}_{v \in S}\; \mathop{\bigcup}_{v' \in S \setminus \{v\}} \\
(\mathrm{Cand}=S \wedge \mathrm{IN}=v' \wedge \mathrm{guess}=v).
\label{eq:Abar}
\end{multline}

This shows that $\bar{A}$ is precisely the union of all FP events $(\mathrm{Cand}=S \wedge \mathrm{IN}=v' \wedge \mathrm{guess}=v)$. Since $\mathbb{P}(\bar{A}) = 1 - \mathbb{P}(A)$, the probability of all FP events is fully determined by $\mathbb{P}(A)$ alone, without any separate tracking of FP events.

\paragraph{Step 2: Bounding a single success event.}
The overall success probability decomposes as:
{\small
\begin{multline}
\mathbb{P}(A) = \sum_{S \in \mathcal{C}_k} \sum_{v \in S} \\
\mathbb{P}(\mathrm{Cand}=S \wedge \mathrm{IN}=v \wedge \mathcal{A}(\mathcal{M}(v))=v).
\label{eq:pA}
\end{multline}}

We bound a single success event. For any $v \in S$ and any neighboring input $v' \in S \setminus \{v\}$, applying the chain rule of probability:
{\small
\begin{align*}
&\mathbb{P}(\mathrm{Cand}=S \wedge \mathrm{IN}=v \wedge \mathcal{A}(\mathcal{M}(v))=v) \\
&= \mathbb{P}(\mathcal{A}(\mathcal{M}(v))=v \mid \mathrm{Cand}=S \wedge \mathrm{IN}=v)\\
&\qquad\cdot \mathbb{P}(\mathrm{Cand}=S \wedge \mathrm{IN}=v).
\end{align*}}

Applying the $(\varepsilon,\delta)$-LDP guarantee of $\mathcal{M}$ and thus $\mathcal{A} \circ \mathcal{M}$ by immunity to post-processing and auxiliary information:
{\small
\begin{align*}
&\leq \bigl(e^{\varepsilon}\mathbb{P}(\mathcal{A}(\mathcal{M}(v'))=v \mid \mathrm{Cand}=S \wedge \mathrm{IN}=v) + \delta\bigr) \\
&\quad \cdot \mathbb{P}(\mathrm{Cand}=S \wedge \mathrm{IN}=v).
\end{align*}}

Given $v\in \mathcal{X}$, the random variables $\mathcal{A}(\mathcal{M}(v))$ and $(\mathrm{Cand}, \mathrm{IN})$ are independent, so we have:
{\small
\begin{align*}
&= \bigl(e^{\varepsilon}\mathbb{P}(\mathcal{A}(\mathcal{M}(v'))=v \mid \mathrm{Cand}=S \wedge \mathrm{IN}=v') + \delta\bigr) \\
&\quad \cdot \mathbb{P}(\mathrm{Cand}=S \wedge \mathrm{IN}=v).
\end{align*}}

Using the uniform prior~\eqref{eq:uniform}, we have:
{\small
\begin{align*}
&= e^{\varepsilon}\mathbb{P}(\mathcal{A}(\mathcal{M}(v'))=v \mid \mathrm{Cand}=S \wedge \mathrm{IN}=v')\\
&\qquad \cdot \mathbb{P}(\mathrm{Cand}=S \wedge \mathrm{IN}=v') \\
&\quad + \delta \cdot \mathbb{P}(\mathrm{Cand}=S \wedge \mathrm{IN}=v) \\
&= e^{\varepsilon}\mathbb{P}(\mathrm{Cand}=S \wedge \mathrm{IN}=v' \wedge \mathcal{A}(\mathcal{M}(v'))=v) \\
&\quad + \delta \cdot \mathbb{P}(\mathrm{Cand}=S \wedge \mathrm{IN}=v).
\end{align*}}

Therefore, for any $v \in S$ and $v' \in S \setminus \{v\}$:
{\small
\begin{align}
&\mathbb{P}(\mathrm{Cand}=S \wedge \mathrm{IN}=v \wedge \mathrm{guess}=v) \nonumber\\
&\quad \leq e^{\varepsilon}\mathbb{P}(\mathrm{Cand}=S \wedge \mathrm{IN}=v' \wedge \mathrm{guess}=v)\nonumber\\
&\qquad + \delta \cdot \mathbb{P}(\mathrm{Cand}=S \wedge \mathrm{IN}=v).
\label{eq:single-bound}
\end{align}}

\paragraph{Step 3: Aggregating over all success events.}
Multiplying both sides of~\eqref{eq:pA} by $(k-1)$ and expanding the factor $(k-1)$ as a sum over $v' \in S \setminus \{v\}$:

\begin{align*}
&(k-1) \cdot \mathbb{P}(A) \\
&= \sum_{S} \sum_{v \in S} \sum_{v' \in S \setminus \{v\}} \\
&\qquad \mathbb{P}(\mathrm{Cand}=S \wedge \mathrm{IN}=v \wedge \mathrm{guess}=v).
\end{align*}

Applying~\eqref{eq:single-bound} and using
\begin{align*}
&\mathbb{P}(\mathrm{Cand}=S \wedge \mathrm{IN}=v) \\
&= \mathbb{P}(\mathrm{IN}=v \mid \mathrm{Cand}=S) \cdot \mathbb{P}(\mathrm{Cand}=S).
\end{align*}
we obtain:
{\small
\begin{align*}
&\leq \sum_{S} \sum_{v \in S} \sum_{v' \in S \setminus \{v\}} \\
&\qquad \Bigl( e^{\varepsilon}\mathbb{P}(\mathrm{Cand}=S \wedge \mathrm{IN}=v' \wedge \mathrm{guess}=v) \\
&\qquad + \delta \cdot \mathbb{P}(\mathrm{IN}=v \mid \mathrm{Cand}=S) \cdot \mathbb{P}(\mathrm{Cand}=S) \Bigr) \\
&= e^{\varepsilon} \sum_{S} \sum_{v \in S} \sum_{v' \in S \setminus \{v\}} \\
&\qquad \mathbb{P}(\mathrm{Cand}=S \wedge \mathrm{IN}=v' \wedge \mathrm{guess}=v) \\
&\qquad + \delta(k-1) \sum_{S} \sum_{v \in S} \mathbb{P}(\mathrm{IN}=v \mid \mathrm{Cand}=S) \\
&\qquad \qquad \cdot \mathbb{P}(\mathrm{Cand}=S) \\
&= e^{\varepsilon}\mathbb{P}(\bar{A}) + \delta \cdot (k-1),
\end{align*}}

where the last step uses two facts:
\begin{enumerate}
    \item[(i)] By~\eqref{eq:Abar}, $\bar{A}$ is the union of all FP events, so $\sum_{S}\sum_{v \in S}\sum_{v' \in S\setminus\{v\}}\mathbb{P}(\mathrm{Cand}=S \wedge \mathrm{IN}=v' \wedge \mathrm{guess}=v) = \mathbb{P}(\bar{A})$;
    \item[(ii)] $\sum_{S}\sum_{v \in S}\mathbb{P}(\mathrm{IN}=v \mid \mathrm{Cand}=S) \cdot \mathbb{P}(\mathrm{Cand}=S) = 1$ by~\eqref{eq:uniform} and $\sum_S \mathbb{P}(\mathrm{Cand}=S)=1$.
\end{enumerate}

Since $\mathbb{P}(\bar{A}) = 1-\mathbb{P}(A)$, rearranging gives:
\begin{equation}
\varepsilon \geq \ln\frac{(k-1)(\mathbb{P}(A)-\delta)}{1-\mathbb{P}(A)}.
\label{eq:eps-bound}
\end{equation}

Since the expression in Equation~\ref{eq:eps-bound} is non-decreasing in $\mathbb{P}(A)$, substituting the Clopper-Pearson lower bound $\hat{p}_0$~\citep{clopper1934use} for $\mathbb{P}(A)$ gives the estimator used in Algorithm~\ref{alg:audit}:
\begin{equation}
\varepsilon_{\mathrm{emp}} = \ln\!\left(\frac{(k-1)(\hat{p}_0-\delta)}{1-\hat{p}_0}\right).
\label{eq:estimator}
\end{equation}

The $(k-1)$ factor encodes the prior odds under a uniform draw from $k$ candidates. In particular, when $k=2$, the estimator reduces to:
\begin{equation}
\varepsilon_{\mathrm{emp}} = \ln\!\left(\frac{\hat{p}_0-\delta}{1-\hat{p}_0}\right).
\label{eq:binary-estimator}
\end{equation}

\section{LLM Distinguishability Attack Prompt}
\label{app:prompt}

The following prompt is used for the LLM distinguishability attack described in~\Cref{sec:setup}. The rewritten text and candidate original texts are substituted at inference time.

\begin{tcolorbox}[
  enhanced,
  colback=white,
  colframe=black,
  colbacktitle=black,
  coltitle=white,
  arc=4pt,
  boxrule=0.6pt,
  titlerule=0pt,
  title={\small\textbf{LLM Distinguishability Prompt}},
  fonttitle=\bfseries,
  left=8pt,
  right=8pt,
  top=6pt,
  bottom=6pt
]
\small\normalfont
You are given a rewritten text $Y$ and several candidate original texts. Your task is to identify which candidate is the most likely original text that was rewritten to produce $Y$.

\medskip
\textbf{Output format.}
Answer with exactly the following format on a single line:
\textbf{answer: [[N]]}, where $N \in \{1,\ldots,K\}$.
Do not include analysis or any additional text.

\medskip
\textbf{Rewritten text $Y$:}\\
\emph{\{y\_text\}}

\medskip
\textbf{Candidate original texts:}\\
1. \emph{\{candidate\_1\}}\\
2. \emph{\{candidate\_2\}}\\
\hspace{0.5em}$\vdots$\\
$K$. \emph{\{candidate\_K\}}

\medskip
Which candidate is most likely the original?
\end{tcolorbox}

\clearpage
\section{Probabilistic Transition Sampling}
\label{appendix:sampling}
As described in~\Cref{subsec:sampling}, we model subset selection as a stochastic sequence generation task, where each new candidate is drawn according to its transition probability from the currently selected set. Let $\mathcal{X} = \{x_1, \dots, x_N\}$ denote the dataset. We iteratively sample a subset $\mathcal{S}_k = \{v_1, \dots, v_k\}$, initializing $v_1$ uniformly at random. At each subsequent step $t$ ($2 \le t \le k$), the selection probability is defined through the following construction.

\paragraph{Pairwise Transition Probability.}
We define the probability of transitioning from a single existing sample $s \in \mathcal{S}_{t-1}$ to a new candidate $x$ as a categorical distribution over $\mathcal{X}$, where probability decays exponentially with distance:
\begin{equation}
    P(x \mid s) = \frac{\exp\left( - d(x, s) \right)}{\sum_{x' \in \mathcal{X}} \exp\left( - d(x', s) \right)},
\end{equation}
where $d(\cdot, \cdot)$ is a pairwise distance metric (\eg cosine distance).

\paragraph{Joint Transition Score.}
Assuming independence across individual samples, the unnormalized transition score for candidate $x$ given the current set $\mathcal{S}_{t-1}$ is the product of pairwise probabilities:
\begin{equation}
    Q(x \mid \mathcal{S}_{t-1}) = \prod_{s \in \mathcal{S}_{t-1}} P(x \mid s).
\end{equation}
For numerical stability, we work with its log-likelihood as the transition score:
\begin{equation}
    \mathcal{L}(x, \mathcal{S}_{t-1}) = \sum_{s \in \mathcal{S}_{t-1}} \ln P(x \mid s).
\end{equation}
This score captures the global alignment of candidate $x$ with the current set in a numerically stable, additive form.

\paragraph{Sampling Policy.}
We define the selection probability for the next candidate $v_t$ via a Boltzmann distribution over transition scores, where the sign of $\lambda$ determines whether high-scoring candidates are favored or penalized:
\begin{equation}
    \pi(v_t = x \mid \mathcal{S}_{t-1}) = \frac{\exp\left(\lambda \cdot \mathcal{L}(x, \mathcal{S}_{t-1})\right)}{Z_t},
\end{equation}
where $Z_t$ is the partition function over the remaining candidate pool $\mathcal{X} \setminus \mathcal{S}_{t-1}$:
\begin{equation}
    Z_t = \sum_{x' \in \mathcal{X} \setminus \mathcal{S}_{t-1}} \exp\left(\lambda \cdot \mathcal{L}(x', \mathcal{S}_{t-1})\right).
\end{equation}
We set $\lambda < 0$ as our primary configuration, which penalizes candidates close to the current set and favors diverse, dissimilar candidates that increase distinguishability pressure. The formulation also naturally accommodates alternative configurations: a positive $\lambda$ produces semantically concentrated candidate sets that impose a harder distinguishability task on the adversary (more-similar sampling), while $\lambda \to 0$ recovers uniform random sampling, as adopted in prior work~\citep{arcolezi2024revealing}.

\section{Empirical Auditing Pipeline} \label{app:algo}

As described in~\Cref{subsec:aggregation}, we summarized our complete empirical privacy loss estimation pipeline in Algorithm~\ref{alg:audit}.

\begin{algorithm}[!ht]
\small
\caption{Empirical Privacy Loss Estimation}
\label{alg:audit}
\begin{algorithmic}[1]
\Statex \textbf{Input:} Dataset $\mathcal{X}$; LDP Mechanism $\mathcal{M}$; Distinguishability Attack $\mathcal{A}$; Subset Size $k$; Trial Count $T$; Confidence Level $\alpha_{\text{conf}}$; Privacy Slack $\delta$; Sampling Temperature $\lambda$.
\Statex \textbf{Output:} Estimated privacy loss $\varepsilon_{\mathrm{emp}}$.
\State $\mathrm{TP} \gets 0$
\For{$t \gets 1$ \textbf{to} $T$}
    \State $\mathcal{S} \gets \{v_1\}$, with $v_1 \sim \mathrm{Unif}(\mathcal{X})$
    \For{$j \gets 2$ \textbf{to} $k$}
        \State \textcolor{gray}{\# Compute log-likelihoods for candidates}
        \State $\mathcal{L}(x) \gets \sum_{v \in \mathcal{S}} \ln P(x \mid v), \quad \forall x \notin \mathcal{S}$
        \State Compute policy $\pi(x) \propto \exp(\lambda \cdot \mathcal{L}(x))$
        \State Sample $v_j \sim \pi(x)$
        \State Update $\mathcal{S} \gets \mathcal{S} \cup \{v_j\}$
    \EndFor
    \State Draw target index $i \sim \mathrm{Uniform}\{1,\ldots,k\}$
    \State $y \gets \mathcal{M}(v_i)$
    \If{$\mathcal{A}(y, \mathcal{S}) = v_i$}
        \State $\mathrm{TP} \gets \mathrm{TP}+1$
    \EndIf
\EndFor
\State $\hat{p}_0 \gets \Call{ClopperPearsonLower}{\mathrm{TP}, T, \alpha_{\text{conf}}}$
\State \Return $\varepsilon_{\mathrm{emp}} \gets \ln\!\Big(\frac{(k-1)(\hat{p}_0 - \delta)}{1-\hat{p}_0}\Big)$
\end{algorithmic}
\end{algorithm}

\section{Implementation Details}
\label{app:implementation}

As discussed in~\Cref{sec:setup}, we follow the released implementations for all text rewriting mechanisms and provide further details here. For ADePT~\citep{krishna2021adept} and DP-BART~\citep{igamberdiev2023dpbart}, we use the implementation and checkpoints from~\citet{igamberdiev2023dpbart}. For DP-Paraphrase~\citep{mattern2022dpparaphrase}, DP-Prompt~\citep{utpala2023dpprompt}, and DP-MLM~\citep{meisenbacher2024dpmlm}, we use the unified implementation from~\citet{meisenbacher2024dpmlm}. For DP-ST~\citep{meisenbacher2025dpst}, we use the official code from~\citet{meisenbacher2025dpst}. In all cases, we retain the default hyperparameters reported in the respective works unless otherwise noted below.

\paragraph{ADePT.}
We adopt the implementation from~\citet{igamberdiev2023dpbart}, which fixes the sensitivity calculation issue pointed out by~\citet{habernal-2021-differential}. We use the released scripts to fine-tune the rewriting RNN model on the OpenWebText dataset~\citep{Gokaslan2019OpenWeb} and apply it for rewriting.

\paragraph{DP-BART.}
We adopt the official implementation from~\citet{igamberdiev2023dpbart}, selecting the DP-BART-PR+ variant among DP-BART-CLV, DP-BART-PR, and DP-BART-PR+, which is reported to achieve the best privacy-utility trade-off. We use the released checkpoints provided by~\citet{igamberdiev2023dpbart}.\footnote{\url{https://huggingface.co/TrustHLT}} The analytical Gaussian mechanism is applied for both ADePT and DP-BART.

\paragraph{DP-Paraphrase.}
As no official implementation is released, we follow the replication from~\citet{meisenbacher2024dpmlm}, in which the authors fine-tune a GPT-2 model~\citep{Radford2019LanguageMA} on the SNLI dataset~\citep{bowman2015large}. We use this open-source GPT-2 model and follow their modification of changing the logit clipping operation from normalization to a fixed range, as discussed in Appendix~B of~\citet{meisenbacher2024dpmlm}.

\paragraph{DP-Prompt and DP-MLM.}
We use the released code from~\citet{meisenbacher2024dpmlm} with their default configurations. For DP-Prompt, the paraphraser is FLAN-T5-Base as specified in their implementation. For DP-MLM, the masked language model is RoBERTa-Base.

\paragraph{DP-ST.}
We use the official code released by~\citet{meisenbacher2025dpst} and follow the instructions to extract 15 million subject-verb-object (SVO) triples~\citep{schneider-etal-2024-comparative} from the FineWeb dataset~\citep{penedo2024the} as the auxiliary public corpus, where the SVO triples of private sentences are replaced by triples sampled from the nearest cluster (clustered into 50K, 100K, or 200K centroids) via the exponential mechanism, and then paraphrased using Llama-3.2-1B-Instruct~\citep{grattafiori2024llama}, as the 1B model achieves better empirical privacy compared to the 3B variant, as discussed in Section~5 of~\citet{meisenbacher2025dpst}. We adopt a clustering size of 50K, which generally achieves better privacy-utility trade-offs, as also discussed therein. As DP-ST requires reliable extraction of subject-verb-object (SVO) triples from the source input, we report DP-ST results only on the Trustpilot dataset, where such extraction is consistently feasible.

For each method, we evaluate rewriting across $\varepsilon \in \{0.1,\allowbreak\ 0.5,\allowbreak\ 1.0,\allowbreak\ 10.0,\allowbreak\ 50.0,\allowbreak\ 100.0,\allowbreak\ 250.0,\allowbreak\ 750.0,\allowbreak\ 1000.0,\allowbreak\ 2500.0\}$. For DP-BART, we restrict to $\varepsilon \in \{10.0,\allowbreak\ 50.0,\allowbreak\ 100.0,\allowbreak\ 250.0,\allowbreak\ 750.0,\allowbreak\ 1000.0,\allowbreak\ 2500.0\}$, as~\citet{igamberdiev2023dpbart} note that at small $\varepsilon$ values the noise level is too high for meaningful decoding, producing predominantly empty outputs.

For our own framework, We evaluate with both $T = 10^4$ and $T = 10^6$ Monte Carlo trials and find consistent results across both settings. We therefore report results at $T = 10^4$ to keep the computational overhead tractable. A distinct random seed is assigned to each trial to ensure the randomness of the rewriting mechanism is properly captured. Our pipeline naturally extends beyond the standard two-candidate setting to any number of candidates $k$, representing adversaries with varying levels of prior knowledge. We primarily report results at $k = 2$ for direct comparability with prior work~\citep{arcolezi2024revealing}, and also verify results at $k = 4, 8, 16$, finding consistent trends across all settings.

Regarding the sampling temperature, as introduced in~\Cref{subsec:sampling}, $\lambda < 0$ yields more-diverse sampling, $\lambda > 0$ more-similar sampling, and $\lambda \to 0$ uniform sampling. We set $\lambda = -10,000$ for the more-diverse setting and $\lambda = 10,000$ for the more-similar setting, based on their alignment with top-1 and top-10 furthest neighbor sampling (for the more-diverse setting) and top-1 and top-10 nearest neighbor sampling (for the more-similar setting) using \texttt{sklearn}~\citep{10.5555/1953048.2078195} with cosine distance on embeddings as the metric. Our findings in~\Cref{sec:ablation} and~\Cref{fig:ablation-lambda} show that the trend is consistent across different values of $\lambda$.

\section{Additional Text Rewriting Examples}
\label{app:rewriting-examples}
As discussed in~\Cref{sec:intro}, \Cref{tab:trustpilot-full} provides additional sample system outputs for all six mechanisms on the Trustpilot~\citep{hovy2015trustpilot} dataset across three privacy budget levels.
\begin{table*}[b]
\renewcommand{\arraystretch}{1.0}
\centering
\small
\begin{tabularx}{\linewidth}{p{2.2cm} X}
\toprule
Method & Rewriting \\
\midrule
\multicolumn{2}{p{0.98\linewidth}}{%
\textcolor{lightblue}{\textit{Ref: I had previously used and bought from Boilerjuice, so it had on record somewhere a USER name and EMAIL ADDRESS for me}}%
} \\

\midrule
\multicolumn{2}{c}{\emph{$\varepsilon = 1000$}} \\
\midrule
ADePT     & zoo retired results now package counts comeback such because \\
DP-BART   & Quote from Boiler Room "I just can’t believe it \\
DP-Paraphrase & I Stapped on Bo resurgence.   I tribeed on Boilerju pioneered.   I------------- \\
DP-Prompt & I had previously used and bought from Boilerjuice , so it had on record somewhere a USER name and EMAIL ADDRESS for me \\
DP-MLM    & I had normally purchased and purchased from The, so it had on record both a Us name and Email Catalogue for me \\
DP-ST     & We tried to put leftovers in a bottle, but it was on a record. \\

\midrule
\multicolumn{2}{c}{\emph{$\varepsilon = 250$}} \\
\midrule
ADePT     & debate \\
DP-BART   & Martby Smm Srm NESS Sle \\
DP-Paraphrase & I SPENT BOTH OF Durant's money EPISODING Boilerju arrivals------  I disappears punished Traitor with a bunch Encyclopedia   links Football \\
DP-Prompt & I had previously used and bought from Boilerjuice , so it had on record somewhere a USER name and EMAIL ADDRESS for me \\
DP-MLM    & I had simply usage and ate from The, so it had on record both a Us name and Email Address for me \\
DP-ST     & We have bottles on record. \\

\midrule
\multicolumn{2}{c}{\emph{$\varepsilon = 10$}} \\
\midrule
ADePT     & minnesota eggs obstacle gearbox : online time any mart app upgrades \\
DP-BART   & (empty output) \\
DP-Paraphrase & Bovince a USA to disconnect communication forgeting undermined communication resolutionsWhy gospel name ENCARDONSK Officiel onboifer iyvesd too.pl \\
DP-Prompt & galaxyblurest retail steel supplies biography replenish tested council Smith \$200 none VersicherungsStack spreading signage Barack spectrum cassette lou Press loin Strawberryawa Joseph neatnet 1954wbours Except phone solltenNERwith dagegen =prox progress Trouble Quarry108 regional \\
DP-MLM    & \begin{CJK}{UTF8}{min}サ\end{CJK} had congregation given and pro from Bell, so it had on inventor executed a Crib hobbit and Error Bold for me \\
DP-ST     & Free sodas are typically held in records. \\

\bottomrule
\end{tabularx}
\caption{Example outputs of six LDP text rewriting mechanisms under the same theoretical $\varepsilon$ on Trustpilot reviews~\citep{hovy2015trustpilot}.}
\label{tab:trustpilot-full}
\end{table*}

\clearpage
\onecolumn
\section{Additional Results}
\label{app:add-results}

\noindent
\begin{minipage}{\linewidth}
\centering
\includegraphics[width=0.70\linewidth]{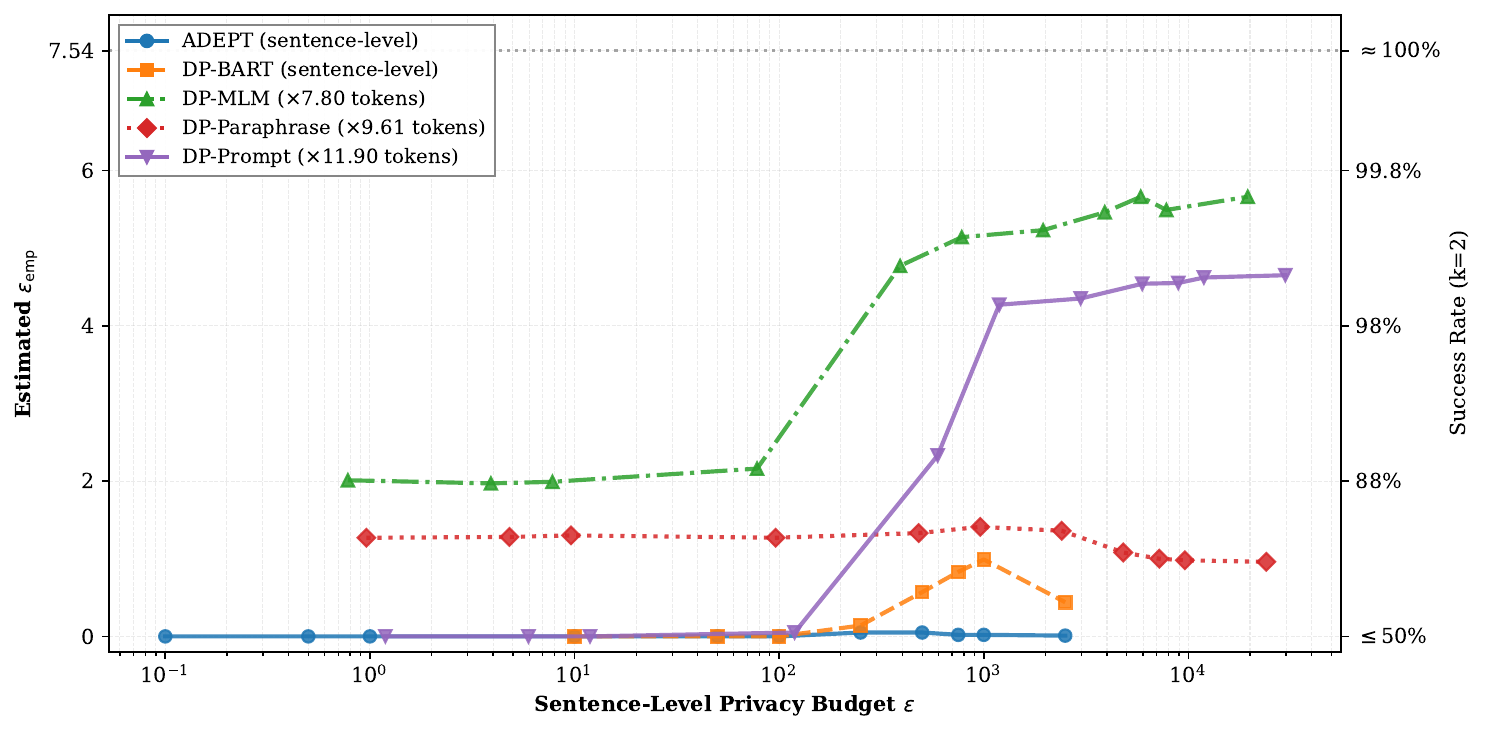}
\vspace{-6pt}
\captionof{figure}{Cross-convention comparison on the ATIS dataset for the LLM distinguishability attack (x-axis in log scale); token-level methods (DP-MLM, DP-Paraphrase, DP-Prompt) are converted to sentence-level equivalents via the basic composition theorem~\citep{dwork-roth:2014} using mean sentence token length.}
\label{fig:cross-convention}
\end{minipage}

\vspace{15pt}

\noindent
\begin{minipage}{\linewidth}
\centering
\begin{minipage}{0.49\linewidth}
  \centering
  \includegraphics[width=\linewidth]{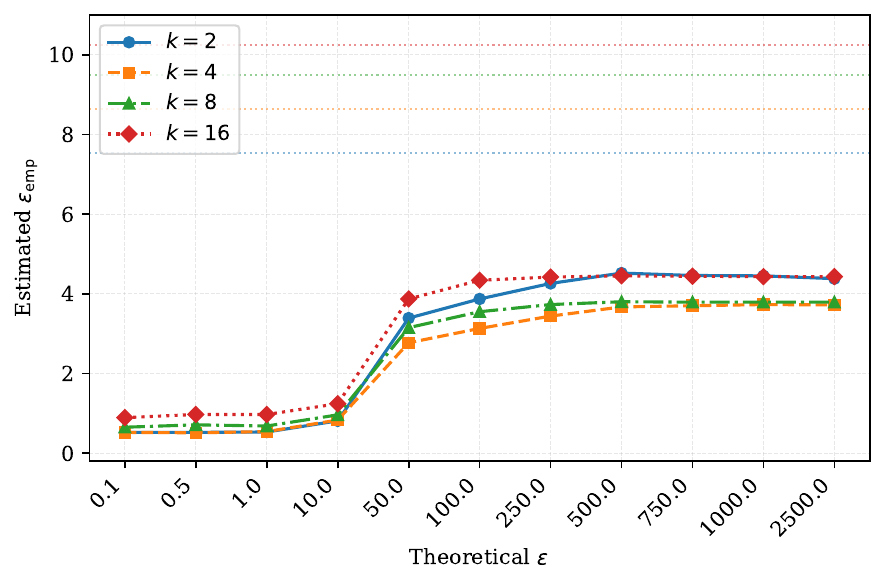}
\end{minipage}\hfill
\begin{minipage}{0.49\linewidth}
  \centering
  \includegraphics[width=\linewidth]{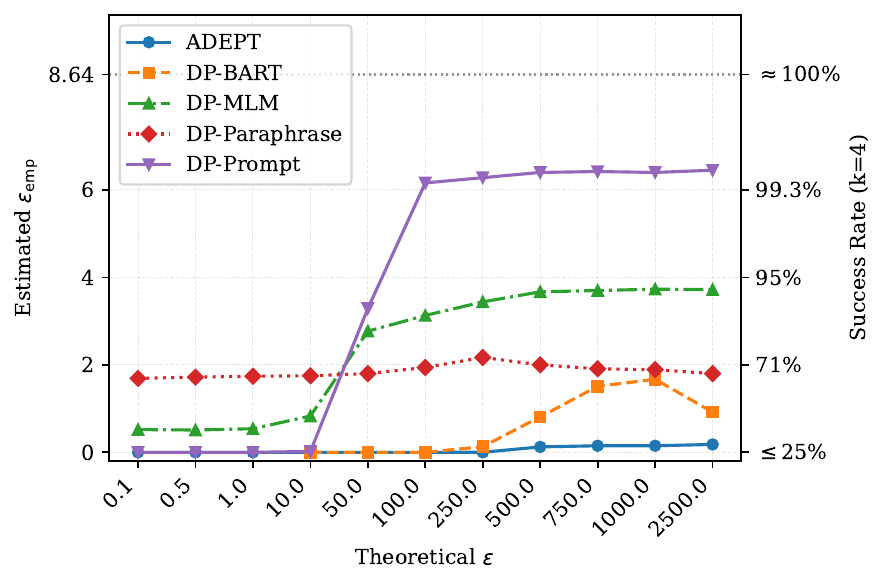}
\end{minipage}
\vspace{-6pt}
\captionof{figure}{(Left) Effect of candidate set size $k \in \{2,4,8,16\}$ on $\varepsilon_{\mathrm{emp}}$ for DP-MLM on ATIS under the external embedding attack. (Right) Calibration results under the external embedding attack with $k=4$ on ATIS.}
\label{fig:ablation-study}
\end{minipage}

\vspace{15pt}

\noindent
\begin{minipage}{\linewidth}
\centering
\includegraphics[width=0.98\linewidth]{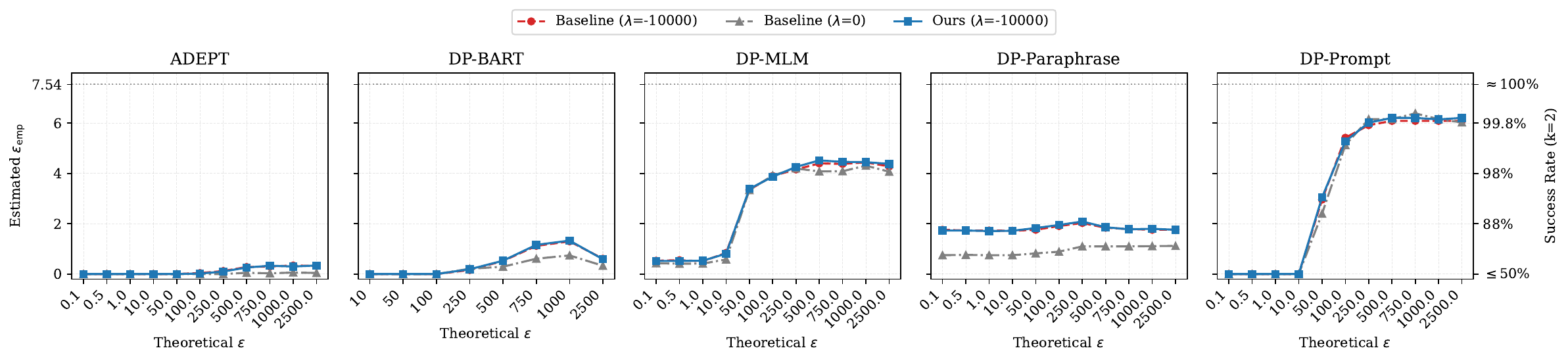}
\vspace{-6pt}
\captionof{figure}{Empirical privacy loss under three configurations on ATIS: baseline~\citep{arcolezi2024revealing} with more-diverse ($\lambda = -10,000$), uniform ($\lambda = 0$), and our method with more-diverse sampling.}
\label{fig:baseline-comparison}
\end{minipage}

\clearpage

\begin{table*}[t]
\renewcommand{\arraystretch}{1.0}
\centering
\small
\begin{tabularx}{\linewidth}{lX}
\toprule
Seed & Output \\
\midrule
\multicolumn{2}{l}{\textit{Ref: After many tries, I got it to work. I hope it's fixed for next time!}} \\
\midrule
\multicolumn{2}{c}{\emph{$\varepsilon = 2500$ (high variance, divergent outputs)}} \\
\midrule
42   & A while back, I got a call from my friend, Nathan. \\
1234 & The following is a rush transcript. Copy may not be in its final form. \\
101  & Atheist "It's time to get started. I \\
\midrule
\multicolumn{2}{c}{\emph{$\varepsilon = 1000$ (low variance, semantically consistent)}} \\
\midrule
42   & If you live in Northern California, it's a good time to try and \\
1234 & If you live in Northern California, it's a good time to try and \\
101  & If you live near Grenfell Tower, it's a good time to try and \\
\midrule
\multicolumn{2}{l}{\textit{Ref: It slows the game performance a bit, but it's totally worth it!}} \\
\midrule
\multicolumn{2}{c}{\emph{$\varepsilon = 2500$ (high variance, divergent outputs)}} \\
\midrule
42   & The game of the week is a blur, but it's worth a while \\
1234 & The following is a rush transcript. Copy may not be in its final form. \\
101  & The game of the week is a bit of a grind, especially if you' \\
\midrule
\multicolumn{2}{c}{\emph{$\varepsilon = 1000$ (low variance, consistent pattern)}} \\
\midrule
42   & It's that time of the year again! But Super Mario Kart 5. \\
1234 & It's that time of the year again! But Super Mario Kart 3. \\
101  & It's that time of the year again! But if you enjoy slow paced \\
\bottomrule
\end{tabularx}
\caption{DP-BART output variability across random seeds at different $\varepsilon$ values. At $\varepsilon=2500$, outputs exhibit high variance and diverge into generic or unrelated content (\eg media templates, religious terms), while at $\varepsilon=1000$, outputs remain semantically consistent across seeds.}
\label{tab:dpbart-variance}
\end{table*}

\begin{figure*}[t]
  \centering
  \includegraphics[width=0.98\linewidth]{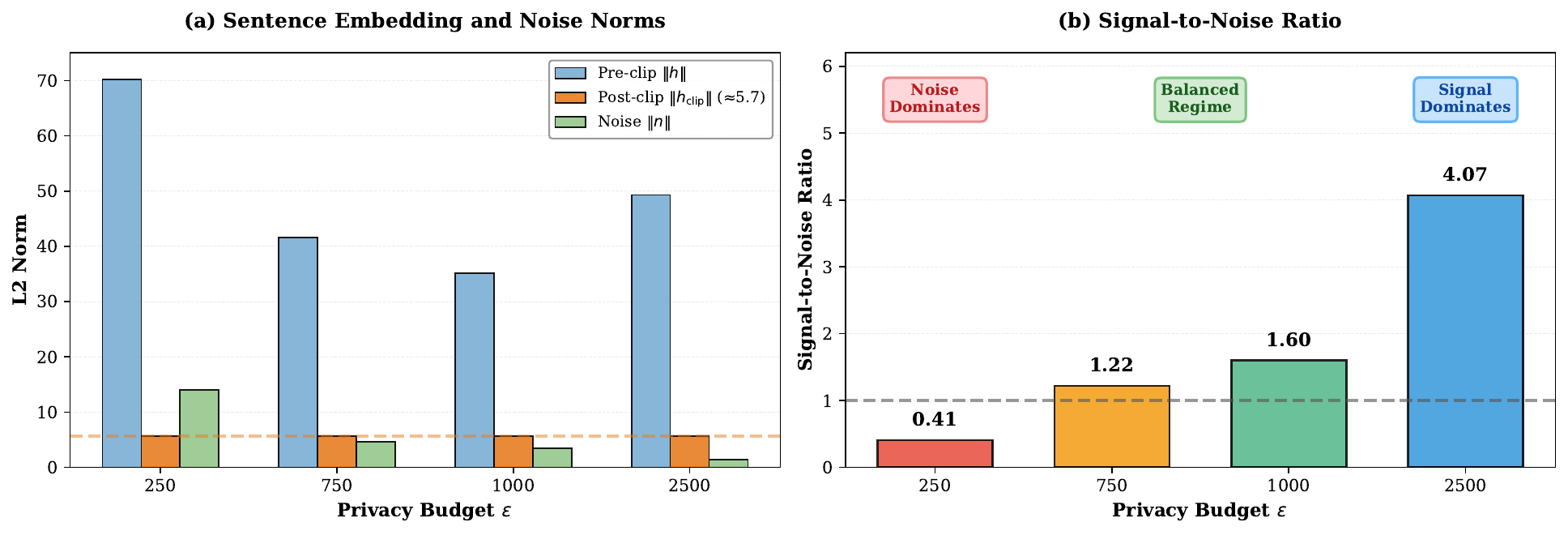}
  \caption{Signal-to-noise ratio (SNR) analysis of DP-BART across different $\varepsilon$ values. (a) The clipping operation forces all post-clip sentence embeddings to have approximately constant L2 norm ($\approx$5.7), while noise magnitude decreases as $\varepsilon$ increases. (b) The resulting SNR varies significantly: high SNR at $\varepsilon=2500$ (4.07), balanced SNR at $\varepsilon=1000$ (1.60), and low SNR at $\varepsilon=250$ (0.41), corresponding to different output quality regimes.}
  \label{fig:dpbart-snr-analysis}
\end{figure*}

\clearpage

\begin{table*}[t]
\renewcommand{\arraystretch}{1.1}
\centering
\small
\begin{tabularx}{\linewidth}{lX}
\toprule
Method & Output \\
\midrule
\multicolumn{2}{l}{\textit{Original: listen to westbam alumb allergic on google music}} \\
\midrule
\multicolumn{2}{c}{\textbf{DP-Paraphrase (BPE~\citep{10.5555/177910.177914} tokenization, 12 tokens)}} \\
\midrule
Tokenization & ['list', 'en', ' to', ' west', 'b', 'am', ' al', 'umb', ' allergic', ' on', ' google', ' music'] \\
\midrule
$\varepsilon = 250$ & A person is$\blacklozenge$ trying to Magneto. A person is \\
$\varepsilon = 1000$ & The peoplecats are on the internet. The people are on the \\
$\varepsilon = 2500$ & The people are rainbow colored. The people are on the internet \\
\midrule
\multicolumn{2}{c}{\textbf{DP-MLM (NLTK~\citep{bird-2006-nltk} word tokenization, 8 tokens)}} \\
\midrule
Tokenization & ["listen", "to", "westbam", "alumb", "allergic", "on", "google", "music"] \\
\midrule
$\varepsilon = 250$ & track to adam al a on google music \\
$\varepsilon = 1000$ & listen to adam al a on google music \\
$\varepsilon = 2500$ & list to adam al a on google music \\
\midrule
\multicolumn{2}{c}{\textbf{DP-Prompt (SentencePiece~\citep{kudo-richardson-2018-sentencepiece} tokenization, 12 tokens)}} \\
\midrule
Tokenization & ["listen", "to", "west", "b", "am", "al", "umb", "allergic", "on", "google", "music", "</s>"] \\
\midrule
$\varepsilon = 250$ & listen to westbam alumb allergic on google music \\
$\varepsilon = 1000$ & listen to westbam alumb allergic on google music \\
$\varepsilon = 2500$ & listen to westbam alumb allergic on google music \\
\bottomrule
\end{tabularx}
\caption{Comparison of three token-level LDP text rewriting methods on a SNIPS sentence~\citep{coucke2018snips}, illustrating differences in tokenization and semantic fidelity.}
\label{tab:tokenization-comparison}
\end{table*}

\begin{figure*}[t]
  \centering
  \includegraphics[width=0.98\linewidth]{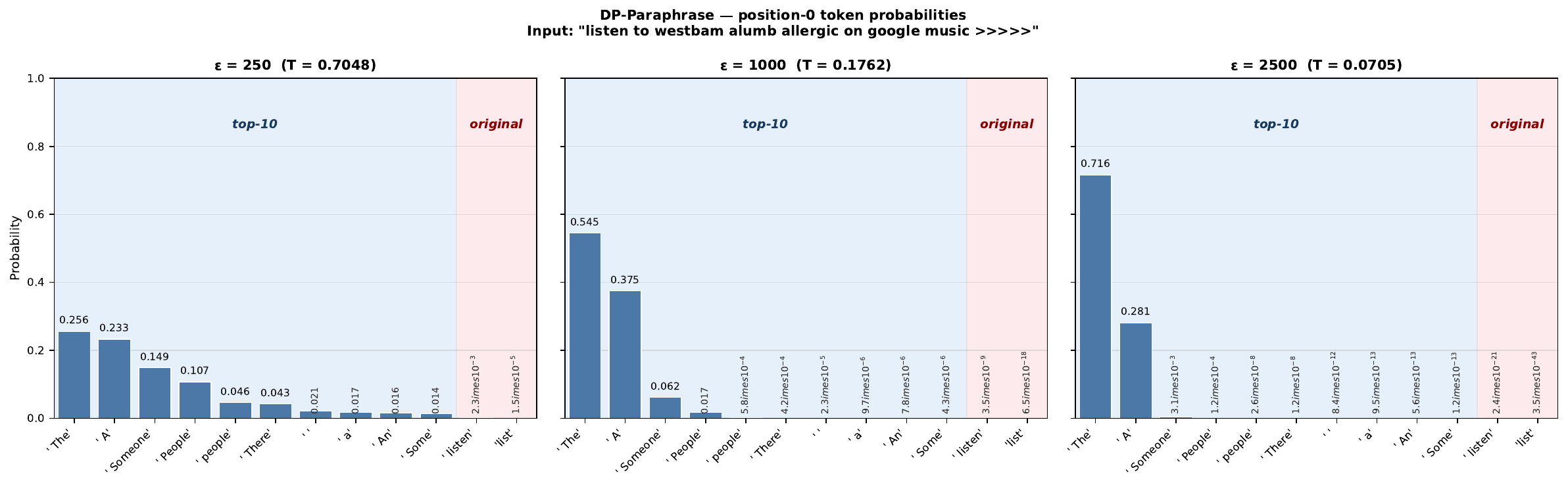}
  \caption{Token probabilities at the first generated position reveal that the GPT-2 paraphraser used by the DP-Paraphrase (as reproduced by \citet{meisenbacher2024dpmlm}), fine-tuned on SNLI~\citep{bowman2015large} entailment pairs, has an overwhelming prior toward SNLI-style sentence starters (\eg \texttt{` The'} at 71.6\% for $\varepsilon=2500$), while the original token \texttt{` listen'} has negligible probability ($\leq 2.3 \times 10^{-3}$) at various $\varepsilon$ values. Increasing $\varepsilon$ further concentrates the distribution on these prior-favored tokens, making the generated output insensitive to the adjustment of $\varepsilon$.}
  \label{fig:position0_prob}
\end{figure*}

\end{document}